\documentclass[longauth]{aa}
\usepackage{graphicx}
\usepackage{txfonts}
\usepackage{hyperref}
\begin{document}
\title{The changing optical and X-ray emission of the dormant $\gamma$\,Cas star HD\,45314}

\author{G.\ Rauw\inst{\ref{ULg}} \corrauth{g.rauw@uliege.be} \and Y.\ Naz\'e\inst{\ref{ULg},\ref{FNRS}} \email{ynaze@uliege.be} \and E.\ Bryssinck\inst{\ref{Bry}} \email{erik.bryssinck@telenet.be} \and W.\ Vollmann\inst{\ref{Vol},\ref{instAAVSO}} \email{vollmann@gmx.at} \and J.\ Guarro Fl\'o\inst{\ref{JGF1},\ref{JGF2}} \email{jngrrfl@gmail.com} \and X.\ Dupont\inst{\ref{instBeSS}} \email{xdupont@wanadoo.fr} \and M.A.\ Smith\inst{\ref{Myron}} \email{myronmeister@gmail.com} \and C.\ Motch\inst{\ref{Motch}} \email{christian.motch@gmail.com} \and \newline G.W.\ Henry\inst{\ref{Hen}} \email{gregory.w.henry@gmail.com} \and J.\ Robrade\inst{\ref{HH} \dag} \email{jan.robrade@uni-hamburg.de} \and A.\ de Bruin\inst{\ref{instBeSS}} \email{ajtdebruin@gmail.com} \and S.\ Charbonnel\inst{\ref{2SPOT}} \email{stephane.charbonnel@2spot.org} \and A.\ Favaro\inst{\ref{Fav}} \email{ andre.favaro@wanadoo.fr} \and P.\ Fricker\inst{\ref{instBeSS}} \email{p.fricker@mpso.ch} \and O.\ Garde\inst{\ref{2SPOT}} \email{olivier.garde@2spot.org} \and O.\ Gayrard\inst{\ref{instBeSS}} \email{o.gayrard@free.fr} \and \newline S.\ G\"ussregen\inst{\ref{instAAVSO},\ref{instBeSS},\ref{Gue}} \email{stefan@guessregen.de} \and F.\ Houpert\inst{\ref{Hou}} \email{franckhoupert@sfr.fr} \and M.\ Larsson\inst{\ref{Lar}} \email{magnus.larsson@saaf.se} \and R.\ Leadbeater\inst{\ref{Lea}} \email{robin@threehillsobservatory.co.uk} \and V.\ Lecocq\inst{\ref{Lec}} \email{ vincentlecocq3@gmail.com} \and P.\ Le D\^u\inst{\ref{2SPOT}} \email{pascal.ledu@2spot.org} \and T.\ Lester\inst{\ref{Les}} \email{t2lester@gmail.com} \and L.\ Mulato\inst{\ref{2SPOT}} \email{lionelmulato@gmail.com} \and \newline T.\ Petit\inst{\ref{2SPOT}} \email{thomas.petit@2spot.org} \and R.\ Pomillo\inst{\ref{instAAVSO}} \email{rosario.pomillo@fastwebnet.it} \and O.\ Thizy\inst{\ref{Thi}} \email{thizy@free.fr} \and F.\ Weil\inst{\ref{instBeSS}} \email{franck.weil@wanadoo.fr}}

\institute{Space sciences, Technologies and Astrophysics Research (STAR) Institute, Universit\'e de Li\`ege, All\'ee du 6 Ao\^ut, 19c, B\^at B5c, 4000 Li\`ege, Belgium \label{ULg} \and Senior Research Associate FRS-FNRS (Belgium) \label{FNRS} \and BRIXIIS Observatory, Eyckensbeekstraat 2, 9150 Kruibeke, Belgium \label{Bry} \and BAV Bundesdeutsche Arbeitsgemeinschaft Veränderliche Sterne \label{Vol} \and Observer of the American Association of Variable Stars Observers (AAVSO) \label{instAAVSO}  \and Piera Remote Observatory, C/.\ J.\ Balmes 2, 08784 Piera, Catalonia, Spain \label{JGF1} \and SMM Remote Observatory, Av.\ de Catalunya 38, 25354 Santa Maria de Montmagastrell, Catalonia, Spain \label{JGF2} \and Observer of the Be Star Spectra (BeSS) database \label{instBeSS} \and NSF OIR Lab, 950 N.\ Cherry Ave., Tucson, AZ 85721, USA \label{Myron} \and Universit\'e de Strasbourg, CNRS, Observatoire Astronomique de Strasbourg, UMR 7550, 67000 Strasbourg, France \label{Motch} \and Tennessee State University (retired), Nashville, TN 37209, USA \label{Hen} \and Hamburger Sternwarte, Gojenbergsweg 112, 21029 Hamburg, Germany \label{HH}  \newline \dag Deceased \and 2SPOT, 45, Chemin du Lac, 38690 Ch\^abons, France \label{2SPOT} \and Bd. Carnot, 21000 Dijon, France \label{Fav} \and Astronomische Arbeitsgemeinschaft Rheingau (AAR), 65385 R\"udesheim am Rhein, Germany \label{Gue} \and Verny Observatory (FR57), 57420 Verny, France \label{Hou} \and Swedish Association of Amateur Astronomers, Sweden \label{Lar} \and Three Hills Observatory, The Birches CA7 1JF, United Kingdom \label{Lea} \and Route du Paradis, 69530 Orlienas, France \label{Lec} \and 1178 Mill Ridge Road, Arnprior, ON, K7S3G8, Canada \label{Les}  \and Observatoire Belle Etoile, 38420 Revel, France \label{Thi}}
\abstract{$\gamma$\,Cas stars are Oe/Be stars that exhibit bright and hard X-ray emission. HD\,45314 belonged to this category, but lost its $\gamma$\,Cas characteristics as its circumstellar disc started to fade away.}{The successive steps of the disc dissipation and its subsequent ongoing rebuilding provide insight into the physical processes at work.}{The star was monitored in optical spectroscopy, optical photometry, and X-ray spectroscopy to follow its variability on timescales of hours to years. Emission line properties were measured on the spectra and compared with optical brightness and colours. Time series of magnitudes, colours, line equivalent widths and violet over red peak intensity ratios were analysed with Fourier methods to uncover possible periodicities.}{For five years, HD\,45314 exhibited oscillations of its magnitude and line strengths on a timescale of 230\,d. It then declined towards a nearly disc-free stage and is now slowly rebuilding its disc. During these phases, the violet over red peak intensity ratio exhibited a modulation on a timescale of about 1190\,d. Over the whole campaign, prominent short-term photometric variations, notably at a frequency of 3.369\,d$^{-1}$, were probably due to $\beta$\,Cep-type pulsations. The amplitude of these pulsations was stronger during phases of overall brightness changes. In parallel, the X-ray spectrum switched to a low state at the onset of the oscillation phase and has remained in this low state since then. Some residual hard emission is nonetheless observed.}
         {HD\,45314's long-term optical variations bear resemblance with the predictions of smooth particle hydrodynamics calculations, although a close look does reveal some inconsistencies. Unlike some other $\gamma$\,Cas stars, where the X-ray emission remained nearly constant despite major changes of the circumstellar disc, HD\,45314's X-ray properties changed dramatically as the disc begun to dissipate. These differences might arise from a wider orbital separation between the Oe star and its putative white dwarf companion or from the ablation of the disc by the radiation field of the Oe star.}  
\keywords{stars: emission line, Be -- stars: individual (HD\,45314) -- stars: variables: general -- stars: oscillations -- X-rays: stars}
\maketitle
\nolinenumbers

\section{Introduction}
Classical Oe/Be stars are rapidly rotating OB main-sequence or giant stars surrounded by a viscous Keplerian decretion disc \citep{Riv13,Riv26}. Be stars often display variable circumstellar emission, alternating between a (near) disc-free state and a state with overwhelmingly strong emission lines \citep[e.g.][]{Ste09,Naz19,Rau21,Baa23,Lab25}. Beside long-term variations that are mostly related to changes of the disc, Oe/Be stars also frequently display short-term photometric \citep[e.g.][]{Sem18,Bal20,Naz20,Lab21} and spectroscopic variability \citep[e.g.][]{Riv01,Mai03,Naz20b}, which is commonly attributed to non-radial pulsations \citep{Riv26} that could play a role in the ejection of photospheric material into the circumstellar disc.

A subgroup of the earliest Be stars display a remarkable hard and bright X-ray emission \citep[see][]{Smi16,Naz18,Naz23}. These objects are named $\gamma$~Cas stars after their prototype, the B0.5\,IVe star $\gamma$~Cassiopeiae, and represent 12\% of the early-type Be stars \citep{Naz23}. They exhibit $\log{L_{\rm X}}$ in the 0.5 -- 10\,keV energy range of $31.4$ -- $33.2$ and $\log{L_{\rm X}/L_{\rm bol}} > -6.4$. This is about ten times higher than the X-ray emission of normal stars of same spectral type but significantly less than the emission of Be High-Mass X-ray binaries \citep[BeHMXBs,][]{Rei11}. The bulk of the X-ray emission of $\gamma$\,Cas stars arises from a hot (kT $\geq 5$\,keV) thermal optically thin plasma, whereas the vast majority of OB stars display a rather soft thermal X-ray emission with $kT \sim$ 0.2 -- 0.6\,keV and $\log{L_{\rm X}/L_{\rm bol}} \sim -7$ \citep{Naz09}. The origin of the unusual emission of $\gamma$\,Cas stars has been a mystery for five decades. The main competing scenarios were accretion onto a white dwarf (WD) companion \citep[e.g.][and references therein]{Mur86,Tsu18,Tsu23,Toa25} or magnetic reconnection events between small-scale magnetic fields at the Be star surface and a magnetic field rooted in the Be disc \citep[][and references therein]{Smi99,Mot15,Smi16,Smi19}. Recent high-resolution {\it XRISM} X-ray spectroscopy of the 6 -- 7\,keV Fe line complex demonstrated that, at least in the case of $\gamma$\,Cas, the hard X-rays arise close to the companion, thereby favouring accretion onto a magnetic WD \citep{Naz26}.  

Currently, the O9:npe star HD\,45314 \citep[= PZ\,Gem,][]{Sot11} is the $\gamma$\,Cas star with the earliest apparent spectral type \citep{Rau13}. HD\,45314 belongs to the original group of Oe stars introduced by \citet{Con74}, although this classification might be biased towards earlier spectral types as a result of circumstellar He\,{\sc i} emission \citep{Neg04}. In this work, we discuss the considerable variations in optical photometry as well as in optical and X-ray spectroscopy over the past three decades. The star entered a phase of strong variability in 2014 \citep{Rau15,Rau18}, and its circumstellar disc recently experienced a phase of dissipation. This phenomenon is not only interesting for the study of Oe/Be discs, but offers also an ideal opportunity to study the connection between the disc and the $\gamma$\,Cas-like emission. 

\section{Observations}
\subsection{Optical spectroscopy \label{obs:spectro}}
Between 2017 and 2024, we continued our spectroscopic monitoring of HD\,45314 with the 1.2\,m robotic TIGRE telescope \citep{Schmitt,Gon22} at La Luz Observatory near Guanajuato (Mexico). TIGRE was equipped with the refurbished HEROS echelle spectrograph \citep{Kaufer2,Schmitt} providing a resolving power of 20\,000 over the 3760 -- 8700\,\AA\ wavelength range with a small gap around 5600\,\AA. Weather permitting, HD\,45314 was observed every two weeks over the yearly six months visibility season to follow the long-term evolution of its circumstellar emission lines. In December 2019, we further performed an intensive spectroscopic monitoring with TIGRE. During four consecutive nights, we collected 206 spectra each with a 6\,min exposure time. During the first two nights, the star was observed for nearly seven hours in a row. During the third and fourth nights, the star was followed respectively for four and three continuous hours. All TIGRE spectra were reduced and calibrated with the HEROS reduction pipeline \citep{Mittag,Schmitt}.

In September 2020 and October 2021, we used the Aur\'elie spectrograph \citep{Gil94} on the 1.52\,m telescope at Observatoire de Haute Provence (OHP) to collect respectively 38 and 54 spectra. Aur\'elie was equipped with an Andor-Newton CCD camera and grating \#3, offering a resolving power of 11\,000 over the wavelength range from 6430 -- 6780\,\AA. Integration times varied between 15 and 40 minutes, depending on the weather conditions. The observations were spread over five (2020) or six (2021) consecutive nights, and spectra were collected continuously over intervals of typically 2.5 - 3 hours. The same equipment was employed in September 2023 and October 2024 to collect 13 and 6 blue spectra, respectively, covering the 4450 -- 4890\,\AA\ domain at a resolving power of about 7\,000.

These data were complemented by a large set of amateur spectra. The majority were taken with off-the-shelf long-slit spectrographs, focusing on the spectral region around the H$\alpha$ line. Some observers used either commercial or self-made echelle spectrographs, which also gave access to the blue part of the spectrum. Additional information on the equipment is given in Appendix\,\ref{Bess}. All amateur data are available on the Be Star Spectra database \citep[BeSS,][]{Nei11}\footnote{http://basebe.obspm.fr/basebe/}. 

For the spectral regions around the He\,{\sc i} $\lambda$\,5876 and H$\alpha$ lines, we used the {\tt telluric} tool of the {\sc iraf} software with the atlas of telluric lines of \citet{Hinkle} to correct the telluric absorptions. Further data analysis (continuum normalisation, radial velocity (RV) measurements) was performed with the {\sc midas} software. We measured the equivalent widths (EWs) of the H$\beta$, He\,{\sc i} $\lambda$\,5876, and H$\alpha$ lines by integrating the normalised spectra respectively over the wavelength ranges 4850 -- 4875\,\AA, 5865 -- 5885\,\AA, and 6530 -- 6600\,\AA. Moreover, we measured the velocity separation and intensity ratios, $V/R = \frac{{\cal F}_V-{\cal F}_c}{{\cal F}_R-{\cal F}_c}$, between the continuum-subtracted normalised fluxes of the violet and red peaks of double-peaked emission lines (H$\beta$, He\,{\sc i} $\lambda$\,5876, H$\alpha$, and He\,{\sc i} $\lambda$\,6678). For those data that cover the blue spectral range, we further measured the RV by fitting a Gaussian to the He\,{\sc ii} $\lambda$\,4686 absorption line as well as via cross-correlation of the observed spectra in the 4635 -- 4705\,\AA\ domain with a synthetic TLUSTY template \citep{Lan03}. 

\subsection{Optical photometry}
\subsubsection{Space-borne photometry}
The Transiting Exoplanet Survey Satellite \citep[{\it TESS},][]{TESS} observed HD\,45314 on nine occasions between 2018 and 2026. {\it TESS} carries four wide-field cameras, each of them covering a $24^{\circ} \times 24^{\circ}$ field of view. Each individual pixel corresponds to (21\,arcsec)$^2$ on the sky. The waveband of the cameras ranges between 6000\,\AA\, and 1\,$\mu$m. {\it TESS} monitors sky sectors of $24^{\circ} \times 96^{\circ}$ for about 27 consecutive days. HD\,45314 was observed at a 2\,min cadence during Sectors\,6 (December 2018 -- January 2019), 33 (December 2020 -- January 2021), 43 -- 45 (September 2021 -- November 2021), 71 -- 72 (October 2023 -- December 2023), 87 (December 2024 -- January 2025), and with a 158\,s cadence during the 3I/ATLAS observing campaign (Sector\,1751, 15 -- 22 January 2026). The duration of each dataset was about a month, except for the last one which lasted a week.

The light curves, processed with the {\it TESS} pipeline \citep{Jen16}, were downloaded from the Mikulski Archive for Space Telescopes (MAST\footnote{http://mast.stsci.edu/}). These data provide background-corrected aperture photometry (SAP) as well as Pre-search Data Conditioned (PDC) photometry obtained after correcting trends correlated with systematic spacecraft or instrumental effects. The PDC data were kept for this study as they are further corrected for crowding, the limited size of the aperture, and instrumental systematics. For Sector\,1751, the light curve was extracted from image cutouts of 51$\times$51 pixels using aperture photometry performed with the Python package Lightkurve\footnote{https://lightkurve.github.io/lightkurve/ - high quality data were selected using option quality\_bitmask=`hard' in task {\sc search.tesscut}.}. The background mask was defined by pixels with fluxes below the median flux (i.e.\ below the null threshold). The background was estimated in that region using either a principal component analysis with five components or a simple median. The former method was favoured as it yielded light curves with slighly smaller long-term trends. 
The {\it TESS} fluxes were converted into magnitudes using\footnote{https://tess.mit.edu/public/tesstransients/pages/readme.html\#flux-calibration} $-2.5\,\log(flux) + 20.44$. The overall long-term evolution of the {\it TESS} photometry is in excellent agreement with the ground-based $V$-band photometry described below.

HD\,45314 is a well isolated source. Nevertheless, we querried the third {\it Gaia} data release catalog \citep[DR3,][]{DR3} for nearby sources that could possibly contaminate the {\it TESS} photometry. {\it Gaia}-DR3 lists 39 sources within a radius of 1\,arcmin around HD\,45314. The brightest of these neighbours has a $G_{RP}$-band magnitude of 15.27, that is 9\,mag fainter than our target. We can thus safely conclude that the {\it TESS} data are free of significant contamination. This conclusion agrees with the value of the {\it TESS} CROWDSAP parameter, which estimates the fraction of (background corrected) flux in the photometric aperture attributable to the target. This keyword has a value of 0.998 for each of the eight sectors with PDC photometry.

\subsubsection{Dedicated ground-based observations}
$V$-band photometric measurements of HD\,45314 were collected since February 2018 by co-author W.\ Vollmann using a digital single lens reflex camera. The comparison and check stars were respectively HD\,44497 and HD\,45194. Typical uncertainties on individual data points are 0.015\,mag.

Between September 2018 and November 2020, differential $B$ and $V$-band photometry was acquired with the 0.4\,m T3 automated photometric telescope (APT) operated by Tennessee State University at Fairborn Observatory in southern Arizona \citep{Hen95}. T3 was equipped with a single-channel photometer featuring a temperature- and humidity-controlled EMI 9924B photomultiplier tube. The comparison and check stars were respectively HD\,43947 and HD\,42545. The $1\sigma$ dispersions of the differential photometry of the check star were 0.0062\,mag and 0.0069\,mag respectively in the $B$ and $V$ bands.   

\subsubsection{Archival data}
To study long-term photometric variations, we compiled data from several archives. We used the website of the American Association of Variable Star Observers (AAVSO \footnote{https://www.aavso.org/}) to retrieve 383 visual magnitudes collected by amateur observer Tony Markham from the Variable Star Section of the British Astronomical Association \citep{Mar03}. These data were collected between October 1997 and April 2016 and are provided with a precision of 0.1\,mag. We discarded those data that were explicitly identified as being affected by bright sky or clouds. 

$V$-band photometry of HD\,45314, collected in the framework of the All Sky Automated Survey \citep[ASAS-3,][]{Poj02}, was downloaded from the project website\footnote{https://www.astrouw.edu.pl/asas/}. The data were taken between December 2002 and November 2009 with two wide field cameras installed at Las Campanas observatory in Chile. We filtered the data, retaining only those 263 observations with a quality flag A (best quality). The ASAS-3 catalogue provides magnitude measurements performed for five different apertures with diameters ranging from two to six pixels. In our analysis, we focused on the aperture 1 magnitudes as they had the lowest average photometric errors (0.031\,mag).

Finally, $V$ and $I_c$-band measurements were extracted from the website\footnote{http://kws.cetus-net.org/~maehara/VSdata.py} of the Kamogata/Kiso/Kyoto wide-field survey \citep[KWS,][]{Mae14}. The $V$-band observations started in December 2010, whilst the $I_c$ monitoring began in October 2013. The KWS project uses an $f/2$ telephoto lens together with a small CCD camera and provides $V$ and $I_C$ aperture photometry for stars with magnitudes between 5 and 11. In our analysis, we kept only those data with photometric errors $\leq 0.015$\,mag.

\subsection{X-ray data \label{Xraydata}}
In addition to the two {\it XMM-Newton} and one {\it Suzaku} observations previously discussed in \citet{Rau18}, HD\,45314 was observed anew with {\it XMM-Newton} in March 2025 (JD\,2460745.669 at mid exposure). The observation lasted 29\,ks and was performed with the European Photon Imaging Camera (EPIC) instruments in full frame mode and the thick optical blocking filter to prevent optical loading. We processed the data with the Science Analysis System (SAS) software version 21.0.0 and the current calibration files available in January 2026. EPIC spectra of HD\,45314 were extracted over a 20\,\arcsec\ radius circular region centred on the {\it Gaia} coordinates of the star. For the subsequent analysis, the spectra were binned to achieve a minimum signal-to-noise ratio of 3 or a maximum oversampling of 5.  

HD\,45314 was also observed with the extended ROentgen Survey with an Imaging Telescope Array ({\it eROSITA}) instrument on the Spectrum-Roentgen-Gamma ({\it SRG}) satellite \citep{Pre21}. The four {\it eROSITA} all-sky surveys were performed between December 2019 and December 2021. These data were previously used by \citet{Naz23}. We used the newest processing (c030) of the data to extract a light curve and a combined spectrum. 

\section{Long-term changes in the  circumstellar environment of HD\,45314 \label{longterm}}
        In this section, we examine the decadal behaviour of optical brightness and emission lines as diagnostics of HD\,45314's circumstellar disc. We identify five different disc activity states, and discuss correlations between various spectroscopic and photometric quantities.
          
\begin{figure}[h]
    \resizebox{8.5cm}{!}{\includegraphics{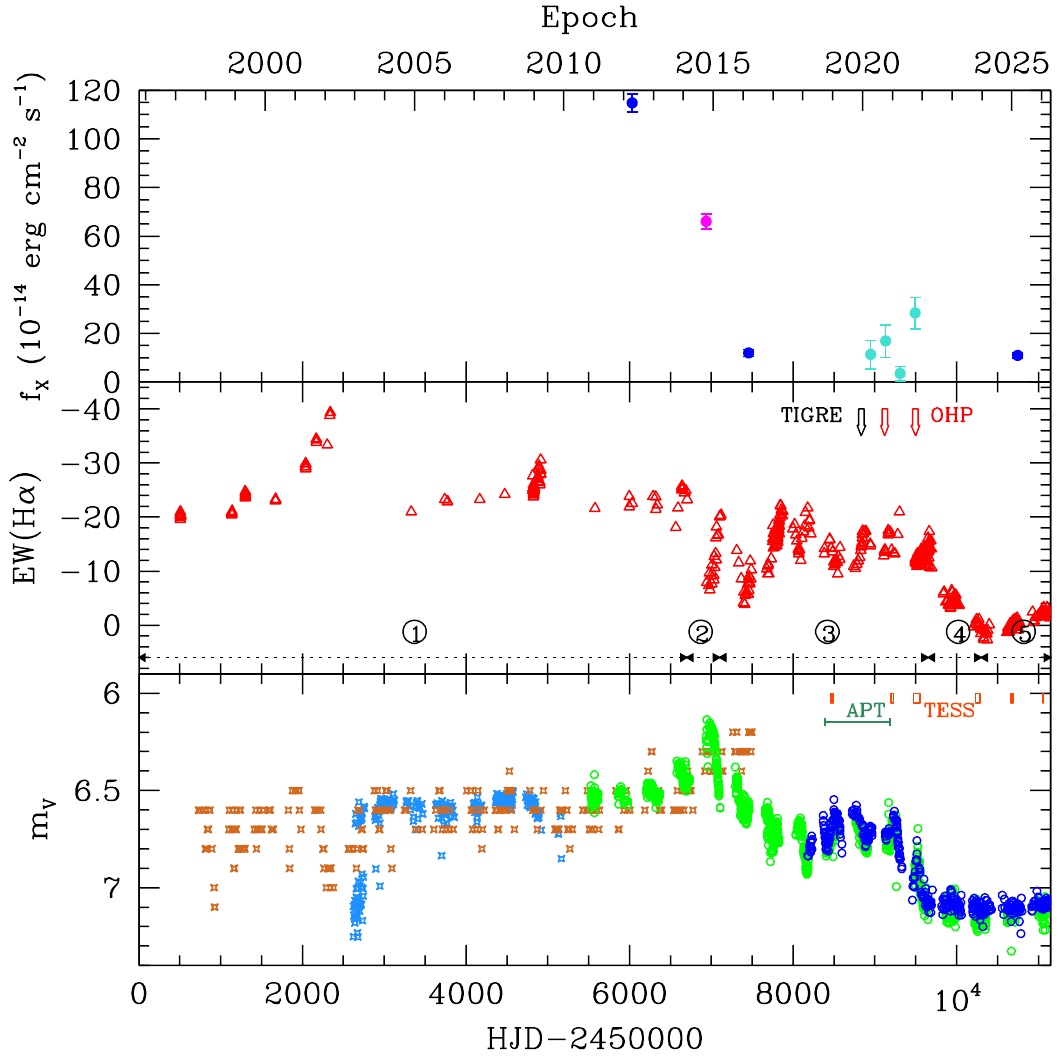}}
    \caption{Historic light curve of HD\,45314 from 1995 until March 2026. The top panel yields the variations in the 0.5 -- 10\,keV observed X-ray flux. Blue, magenta and turquoise symbols stand respectively for {\it XMM-Newton}, {\it Suzaku} and {\it SRG/eROSITA} data. The middle panel displays EW(H$\alpha$) expressed in \AA. The times of the intensive spectroscopic monitoring with TIGRE and with the 1.52\,m OHP telescope are indicated respectively by the black and red arrows. The time intervals highlighted by the dotted line segments at the bottom of the middle panel correspond to the high activity state (1), the shell episode (2), the oscillating phase (3), the decline phase (4), and the low activity and slow recovery state (5). The bottom panel yields the photometric light curve. The symbols stand for visual magnitudes compiled by T.\ Markham (brown crosses), as well as $V$-band magnitudes from ASAS-3 (light blue crosses), from the KWS survey (green circles), and observed by co-author W.\ Vollmann (blue circles). Orange boxes indicate the times of the {\it TESS} observations. The green line highlights the interval of the APT monitoring.\label{historic}}
\end{figure}

Figure\,\ref{historic} illustrates the long-term variations in the $V$-band photometry, EW(H$\alpha$) and X-ray flux of HD\,45314, whilst Fig.\,\ref{historicEW} compares the variations in the EWs of the most prominent emission lines. Since the beginning of our spectroscopic observations, the star displayed a variety of behaviours. We distinguish five different states of circumstellar activity (see Fig.\,\ref{historic}). Between 1995 and 2013, the star was in a high activity state with occasional outbursts. This was followed by a brief shell episode in late 2014. Subsequently, the star entered an oscillating phase with recurrent oscillations both in $m_V$ and EW(H$\alpha$). This lasted until spring 2022 and was followed by a rapid decline in brightness and line emission strength. Finally, since 2024, HD\,45314 is in a low activity state with a slowly recovering circumstellar emission. Some representative H$\alpha$ line profiles are shown in Fig.\,\ref{montage}.     

During the high activity state, the emission lines were well developed and usually displayed a double-peaked profile with some $V/R$ variations. During this high state, in 2003, \citet{Vin09} observed a strong depolarisation effect in the H$\alpha$ line, as expected for a well-developed disc.\footnote{These authors inferred an intrinsic linear polarisation of 0.45\% with a position angle of $135^{\circ}$.} Between 1995 and 2013, the star exhibited episodic outbursts with significant increases, by up to a factor two, in the overall strength of the emission lines (H$\beta$, He\,{\sc i} $\lambda$\,5876, H$\alpha$). The exact behaviour of the various lines during the outbursts changed from one event to another. During the third outburst, which occurred in autumn 2013, towards the end of the high activity state, the three lines displayed narrow single-peaked emissions (see the case of H$\alpha$ in Fig.\,\ref{montage}). Whilst H$\beta$ and He\,{\sc i} $\lambda$\,5876 displayed a clear enhancement of the emission strength, EW(H$\alpha$) underwent a more modest variation during that event. Based on AAVSO and ASAS-3 photometry, \citet{Rau18} reported an anticorrelation between EW(H$\alpha$) and the star's visual brightness. The first {\it XMM-Newton} observation obtained in April 2012, that is during a quiescent phase of the high activity state, displayed the characteristic X-ray spectral energy distribution (SED) of a $\gamma$\,Cas star with a plasma temperature $kT \simeq 15$\,keV, and an Fe emission line complex consisting of a blend of fluorescent Fe K$\alpha$, the Fe\,{\sc xxv} He$\alpha$ triplet and Fe\,{\sc xxvi} Ly$\alpha$ \citep{Rau13,Rau18}. 

At the beginning of the 2014-2015 visibility period, the Be emission lines exhibited a strong narrow absorption core, indicative of a shell-like configuration \citep[][see also Fig.\,\ref{montage}]{Rau18}. These absorption components led to a reduction of the overall strength of the emissions. The lines displayed genuine shell profiles, with narrow absorptions reaching below the continuum, for about two months. Subsequently, the emissions grew again and the profiles progressively returned to those typical of a Be star with an intermediate inclination ($\sim 45^{\circ}$). When referring to HD\,45314's shell episode, we consider the full EW cycle, that is between autumn 2014 and spring 2015, (see Fig.\,\ref{historic}) that included the shell spectra.
The KWS photometry indicates that the brightness in the $V$-band increased during this event. The {\it Suzaku} observation taken during the shell phase (October 2014) again displayed a $\gamma$\,Cas-like X-ray SED, although with a flux half that observed in 2012 \citep{Rau18}. 

The shell event marked the onset of an episode of strong variations, with the emission line EWs undergoing strong oscillations with peak-to-peak amplitudes up to $\sim$ 15\,\AA\ for EW(H$\alpha$) (see Fig.\,\ref{historic}). Until about 2022, the lines remained in emission, although with a progressively decreasing maximum strength. The line morphologies changed in a complicated way, alternating between asymmetric single-peaked, double-peaked and even triple-peaked profiles (see Fig.\,\ref{montageoscill}). The global width of the emission lines, which for steady discs is taken as a proxy of the radial extension of the disc, also considerably changed during this phase (see Fig.\,\ref{montage}).
The He\,{\sc i} $\lambda$\,5876 line displayed a somewhat special behaviour insofar that its maximum strength during these oscillations became occasionally identical to or even exceeded the level seen during the outbursts in the high-state (see Fig.\,\ref{historicEW}). At the beginning of the oscillation phase, in March 2016, a second {\it XMM-Newton} observation unveiled a strong attenuation of the $\gamma$\,Cas characteristics \citep{Rau18}. The overall flux in the 0.5 -- 10\,keV and 2 -- 10\,keV bands was reduced by an order of magnitude compared to March 2012.

Starting from the 2022 -- 2023 observing season, the strength of the emission lines underwent a rapid decline. During most of this phase, the H$\alpha$ emission displayed two emission peaks, again suggesting a disc seen from an intermediate viewing angle. The velocity separation between the peaks increased from $\sim 250$\,km\,s$^{-1}$ in the 2022 -- 2023 season to around $550$\,km\,s$^{-1}$ in 2023 -- 2024, suggesting that the outer disc radius shrank. Simultaneously, H$\alpha$ became dominated by a broad photospheric absorption and EW(H$\alpha$) oscillated between very small negative (emission-dominated) and positive (absorption-dominated) values (see Figs.\,\ref{historic} and \ref{montage}). By end of December 2023, the entire optical spectrum of the star was dominated by broad absorption lines. However, even during that phase, the wings of the H$\alpha$ line exhibited residual emission bumps (see the January 2024 profile in the rightmost panel of Fig.\,\ref{montage}), suggesting that the star was never entirely disc-free. In the subsequent months and years, the emission strength grew slowly and, from January 2025 on, the H$\alpha$ line was again emission-dominated, whilst H$\beta$ remained in absorption. The He\,{\sc i} $\lambda$\,5876 line displayed an intermediate behaviour, going back into emission by the autumn 2025.

\begin{figure}[h]
    \resizebox{8.5cm}{!}{\includegraphics{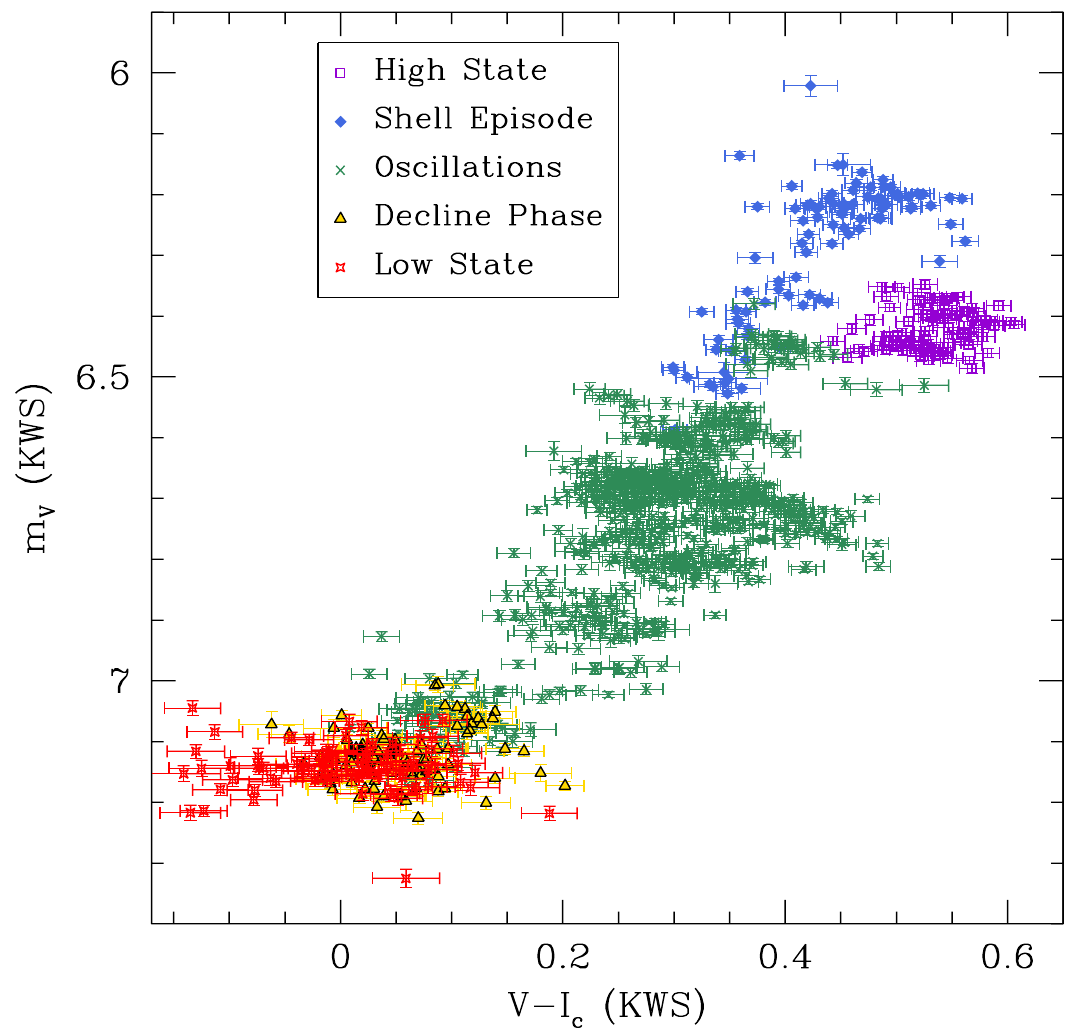}}
    \caption{$V$-band apparent magnitude versus $V-I_C$ colour index as obtained within the KWS survey between autumn 2013 and spring 2026.\label{colour}}
\end{figure}

We used the KWS data to build the time series of the $V-I_c$ colour index of HD\,45314. The photometric decline goes along with the star becoming bluer. Figure\,\ref{colour} illustrates the ($V-I_c$, $m_V$) colour - magnitude diagram with the different states of the circumstellar environment identified by different symbols. On average, the colour and magnitude are well correlated as expected for a circumstellar disc reprocessing stellar light: redder colours are observed when the overall brightness is higher. 

\begin{figure}[h]
    \resizebox{8.5cm}{!}{\includegraphics{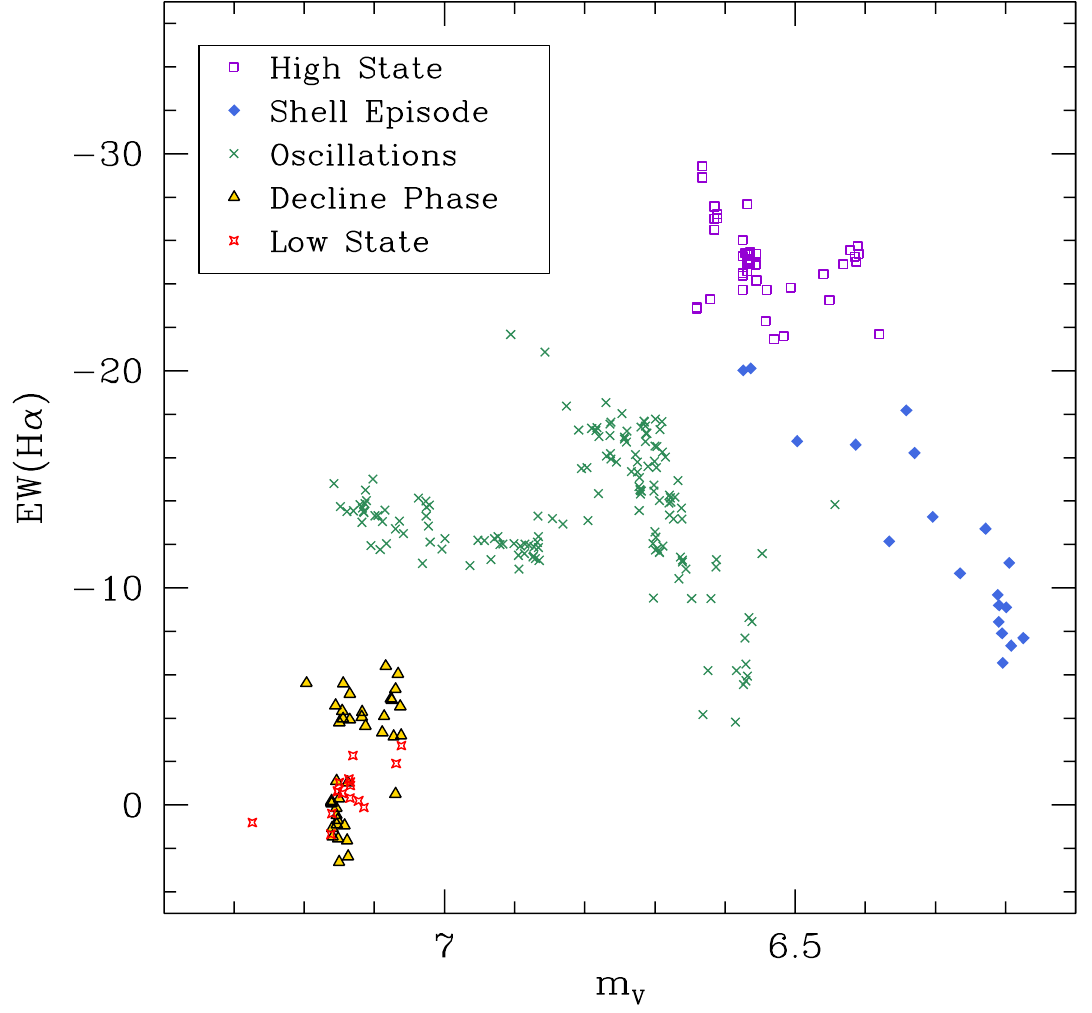}}
    \caption{EW(H$\alpha$) as a function of $V$-band magnitude from the ASAS-3 (high-state only) and KWS (all states) surveys.\label{HavsV}}
\end{figure}

To investigate the link between H$\alpha$ emission strength and optical brightness, we interpolated the $V$-band magnitudes from the nearest-in-time ASAS-3 or KWS data, which fall within less than 15 days from the times of the EW(H$\alpha$) measurements. Figure\,\ref{HavsV} displays the results obtained this way, adopting different symbols for the five different states defined above. HD\,45314 displays a combination of two different behaviours. On the timescale of decades, we observe that the stronger the H$\alpha$ emission (i.e.\ the more negative EW(H$\alpha$)), the brighter the star (i.e.\ the lower $m_V$). The same positive correlation holds for very low levels of circumstellar emission. On timescales of years and for a well-developed disc, we observe instead an anticorrelation with less pronounced H$\alpha$ emission when the star is brighter.

Such an anticorrelation was already found by \citet{Rau18} between the AAVSO visual magnitudes and EW(H$\alpha$) during the high activity state of HD\,45314. A very clear anticorrelation also exists between the KWS $V$-band photometry and EW(H$\alpha$) during the shell episode. The relations of both the high activity state and the shell state appear identical. 

Considering the data of the oscillation phase between 
October 2015 and
February 2021, we observe again an anticorrelation with roughly the same slope as before but occurring around a fainter mean $m_V$ and a weaker H$\alpha$ emission. Starting from October 2021, the relationship shifts towards a fainter $V$-band brightness and a weaker line. At the same time, it also flattens towards a less variable EW(H$\alpha$). Finally, a very different picture holds for the decline phase and the low state where EW(H$\alpha$) displayed a steep positive correlation with $m_V$.

The variations in EW(H$\beta$) and EW(He\,{\sc i} $\lambda$\,5876) are well correlated with EW(H$\alpha$) during the different phases of the disc evolution (see Figs.\,\ref{EWHaHb} and \ref{EWHaHe}). For EW(H$\beta$), the correlations during the different phases have a roughly constant slope, while for EW(He\,{\sc i} $\lambda$\,5876) the slope changes with the state of the disc.

In Oe/Be stars, the H$\alpha$ emission is thought to form over a rather large portion of the disc, whereas the disc's continuum emission arises from the innermost region out to a few stellar radii. The correlation on long timescales thus reflects the variations in the global amount of material in a disc seen under an intermediate inclination. Be stars are thought to gradually lose their disc from the inside outwards. The dissipation of the inner disc can produce a temporary anticorrelation between $m_V$ and EW(H$\alpha$) since the optical brightness decreases faster than the strength of the H$\alpha$ emission, leading to a reduction of the line dilution. 

\section{Medium-term variations \label{mediumterm}}
        In this section, we consider the variability on timescales of several hundred days to a few years that was present during the disc oscillation phase. We search for periodic behaviour among spectrocopic and photometric quantities.
        
\begin{figure}[h]
    \resizebox{8.5cm}{!}{\includegraphics{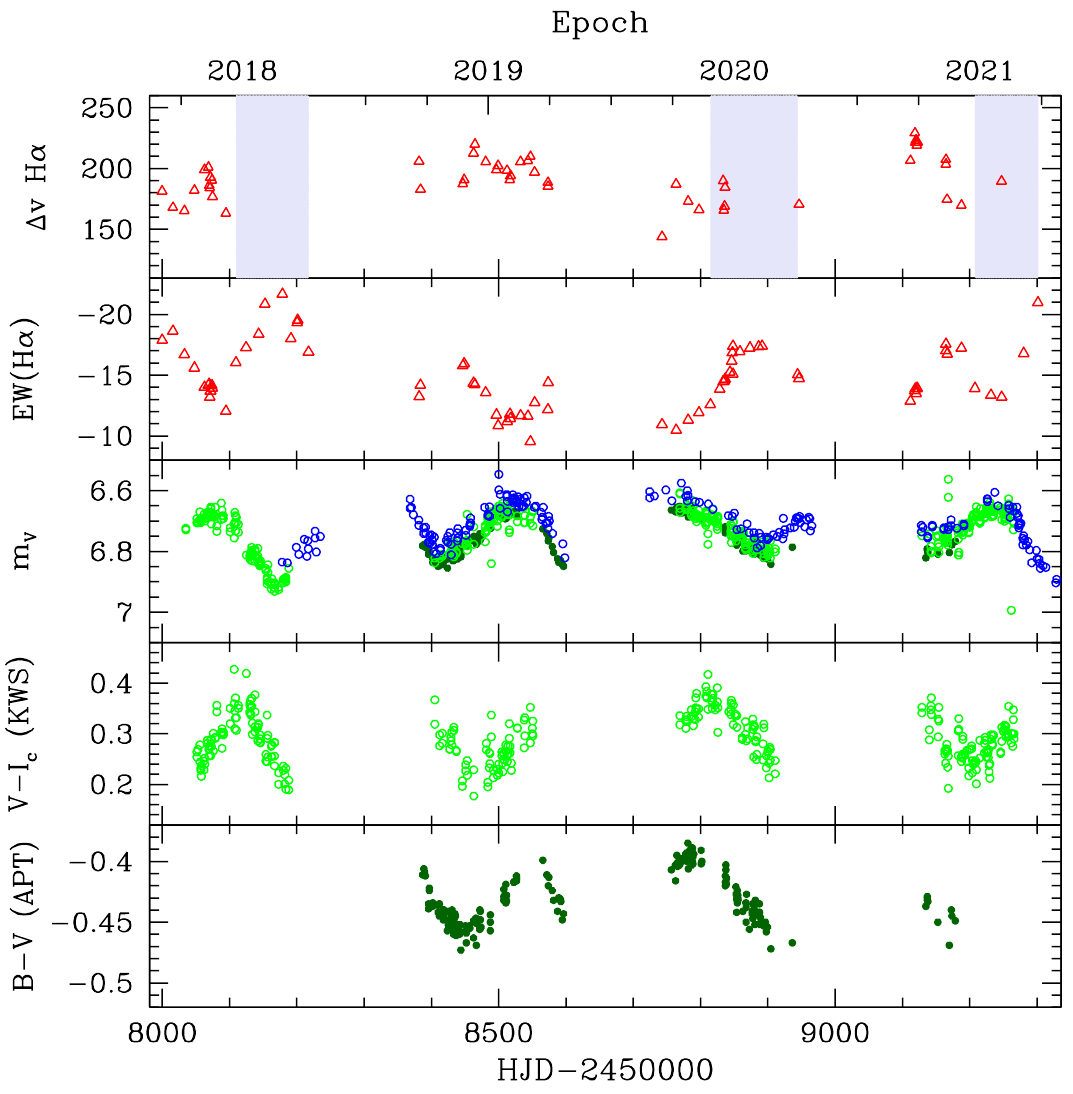}}
    \caption{Variations in peak velocity separation and EW of the H$\alpha$ line (red triangles), in $m_V$, and in colour indices of HD\,45314 during the oscillation phase. Open green circles stand for KWS data, filled dark green circles for APT photometry and open blue circles for measurements by co-author W.\ Vollmann. The shaded areas in the top panel correspond to epochs where H$\alpha$ most of the time displayed only a single peak.\label{oscillations}}
\end{figure}

During the oscillation phase, and more specifically between HJD\,2458000 and 2459400 (autumn 2017 -- spring 2021), the star exhibited cyclic variations in optical brightness, colours, and in the EWs of the emission lines (see Fig.\,\ref{oscillations}). Despite the cyclic variations in EW, we stress that the actual line profiles were quite different between consecutive cycles (see Figure\,\ref{montageoscill}). The velocity separation between the peaks remained nearly constant (see top panel of Fig.\,\ref{oscillations}), though there were episodes where only a single peak could be measured.  Figure\,\ref{oscillations} reveals a time delay between the magnitude and colour variations, with the $B-V$ and $V-I_c$ colour indices lagging behind $m_V$ respectively by about 30 and 50\,days.

\begin{figure}[h]
    \resizebox{8.5cm}{!}{\includegraphics{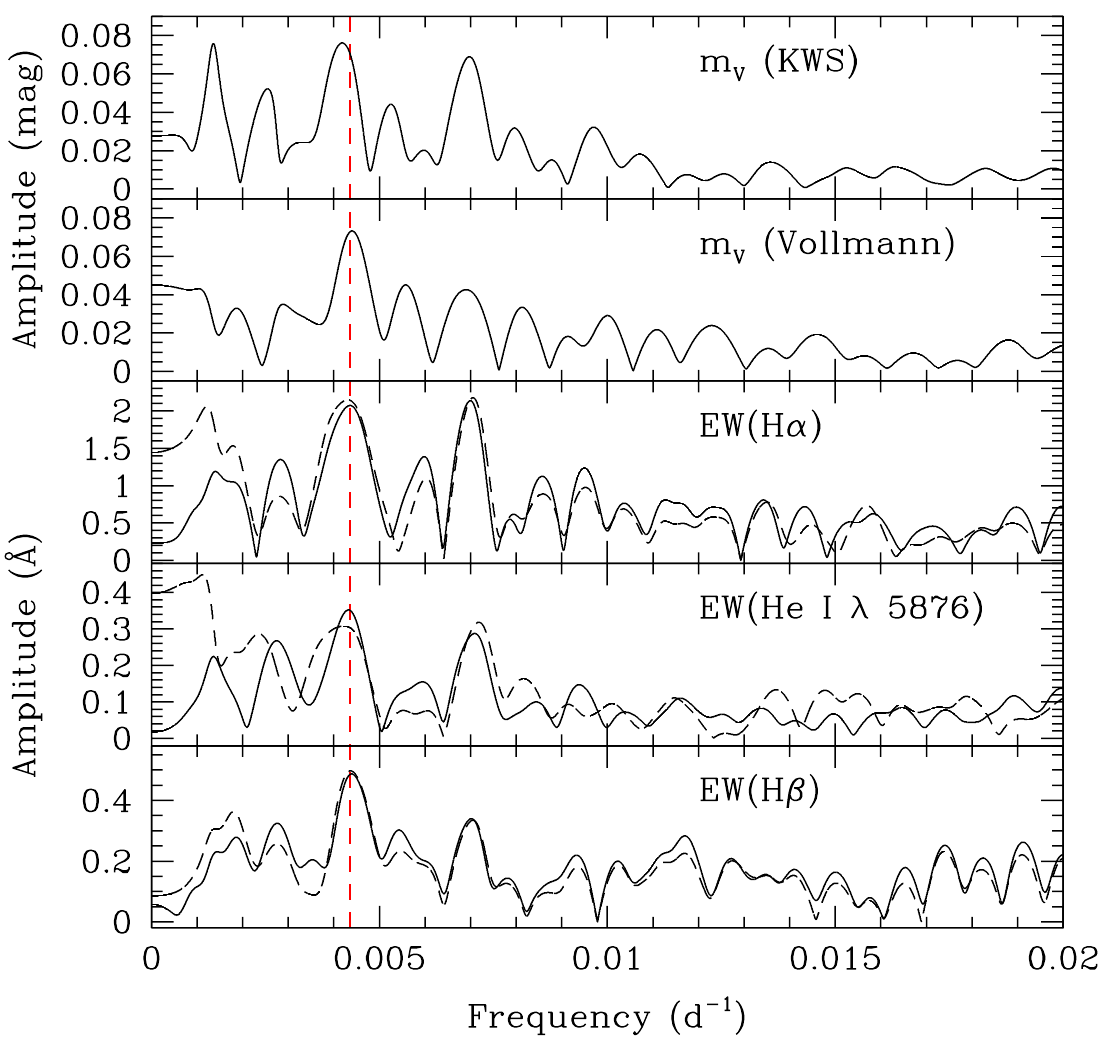}}
    \caption{Fourier periodograms of the $m_V$ magnitudes and EWs of various emission lines in the spectrum of HD\,45314 between HJD\,2458000 and 2459400. The red dashed line indicates a frequency of $4.36 \times 10^{-3}$\,d$^{-1}$. In the three bottom panels, the dashed periodogram corresponds to the raw EW timeseries, whilst the solid line periodogram yields the results after detrending (see text).\label{EW_mv}}
\end{figure}

We performed a Fourier analysis of the various $m_V$ timeseries, as well as of our EW measurements. Throughout this paper, we used the modified Fourier method of \citet{HMM} amended by \citet{Gos01} that explicitly accounts for the irregular sampling of astronomical time series. For the EWs, we performed this analysis also for a detrended timeseries, where we had previously subtracted a long-term trend interpolated from the yearly means. The results of those trials are shown in Fig.\,\ref{EW_mv}. All periodograms exhibit a peak at a frequency close to $4.36 \times 10^{-3}$\,d$^{-1}$, corresponding to a timescale of $230 \pm 4$\,days. A secondary feature near $7.04 \times 10^{-3}$\,d$^{-1}$ corresponds to the yearly alias of the main peak. 

\begin{figure}[h]
    \resizebox{8.5cm}{!}{\includegraphics{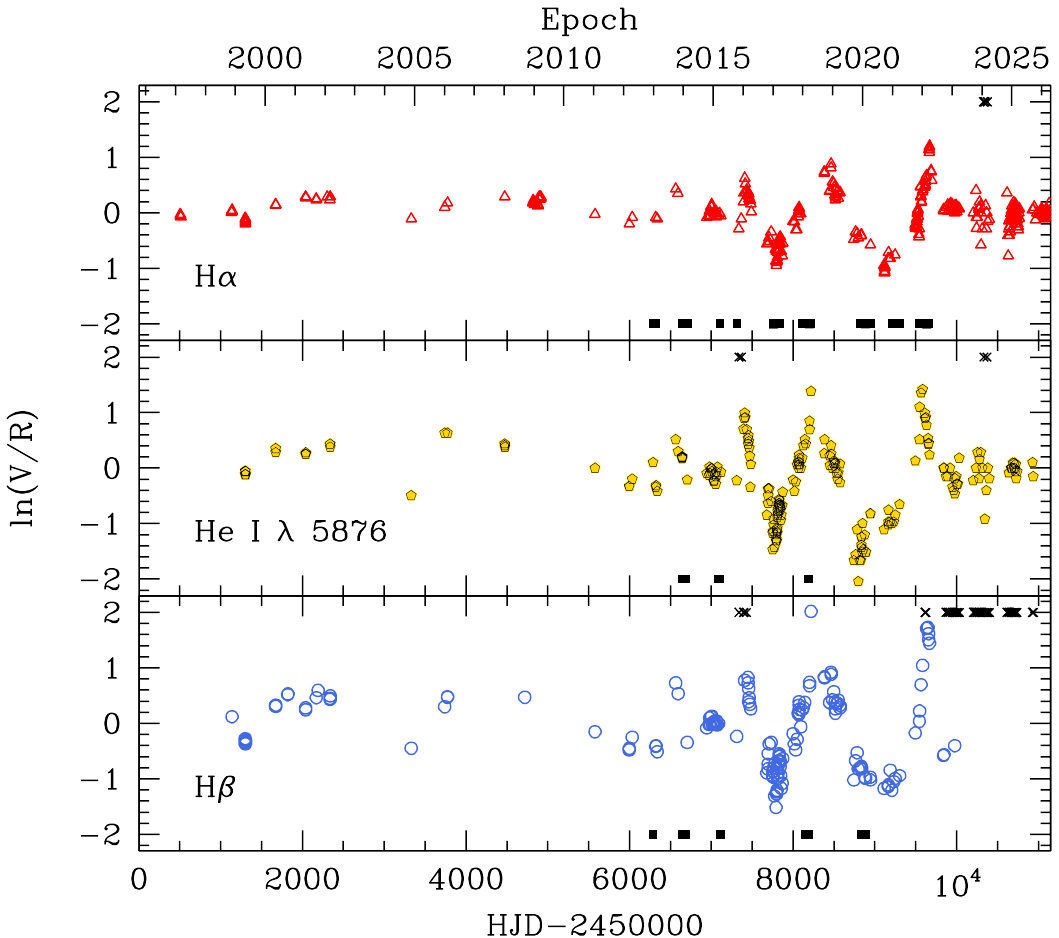}}
    \caption{$\ln(V/R)$ as a function of time over the last three decades. The black squares, arbitrarily put at an ordinate of -2, stand for epochs where the line exhibited a single peak. The crosses, arbitrarily put at an ordinate of +2, indicate observations where V/R was undefined either because of the presence of more than two peaks, or because the line was in absorption. \label{VoR}}
\end{figure}

Variations in the relative strength of the violet and red peaks of prominent emission lines are a common feature of Oe and Be stars. \citet{Rau18} noted the onset of a possible $\sim 1000$\,days cycle in HD\,45314's V/R variations. Our new, more extensive dataset, provides a more detailed view. Figure\,\ref{VoR} displays the value of $\ln(V/R)$ for the three strongest emission lines. A cyclic modulation of V/R is apparent for each of them, but is best seen for H$\beta$ and He\,{\sc i} $\lambda$\,5876. This behaviour started around 2015 and can be traced until 2023. In more recent years, the line profiles were impacted or even dominated by photospheric absorption, preventing us from detecting the cycle after 2023.

\begin{figure}[h]
    \resizebox{8.5cm}{!}{\includegraphics{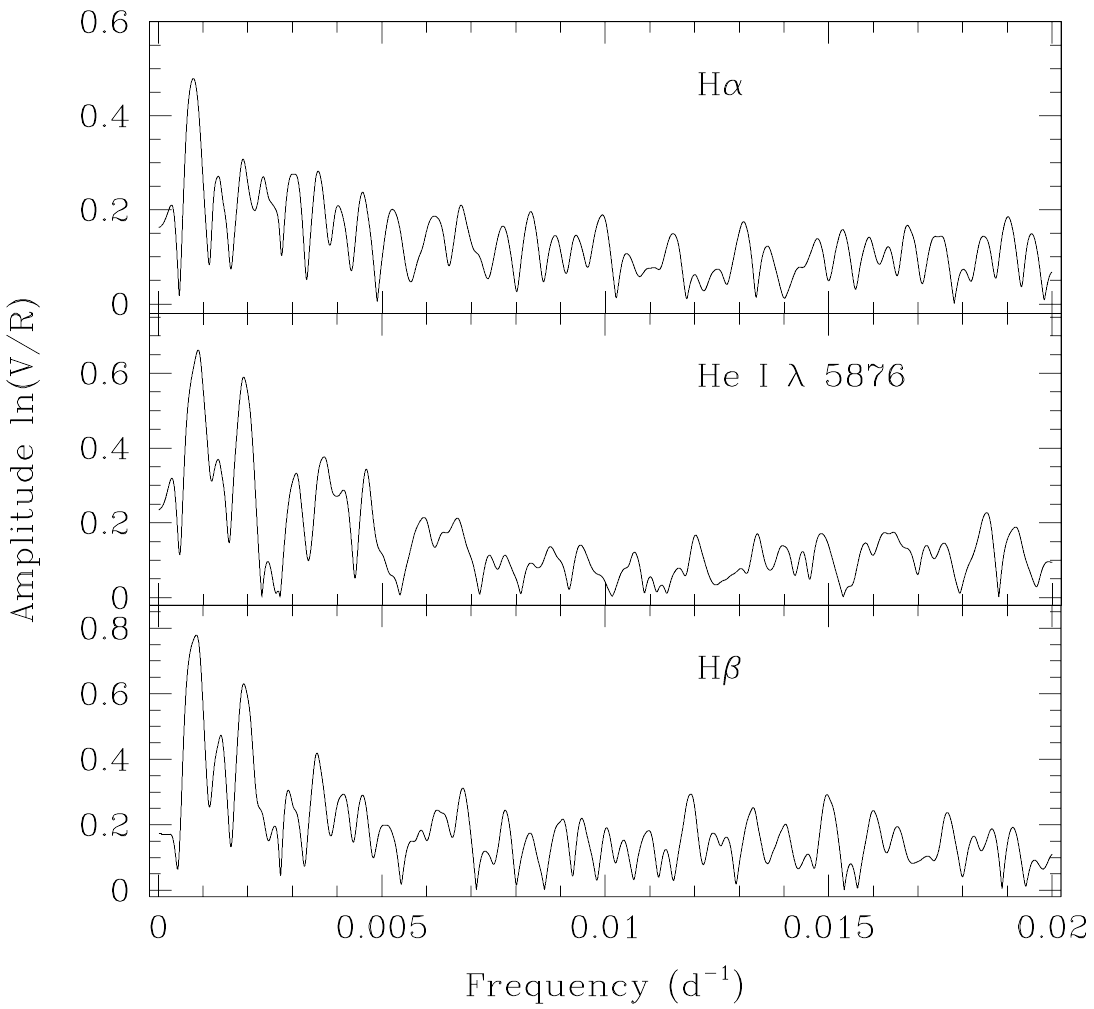}}
    \caption{Fourier periodogram of the time series of $\ln(V/R)$. \label{spVoR}}
\end{figure}

We discarded those data points where V/R was undefined and performed a Fourier analysis of the time series between HJD\,2455990 and 2460400 (see Fig.\,\ref{spVoR}). The highest peaks are found at frequencies of $7.84 \times 10^{-4}$\,d$^{-1}$ (H$\alpha$), $8.52 \times 10^{-4}$\,d$^{-1}$ (H$\beta$) and $8.92 \times 10^{-4}$\,d$^{-1}$ (He\,{\sc i} $\lambda$\,5876). These results point towards a timescale of $1190 \pm 64$\,days, that is $5.17 \pm 0.29$ times the length of the EW and $m_V$ cycles. The Fourier periodograms of the V/R ratios of the H$\beta$ and He\,{\sc i} $\lambda$\,5876 further exhibit a secondary peak at $1.91 \times 10^{-3}$\,d$^{-1}$, corresponding to a timescale of 523.6\,d. Given its frequency, which amounts to 2.14 -- 2.24 times that of the main peak, this secondary peak is unlikely to be a harmonic of the main peak. 

Orbital motion of Be stars in binary systems is commonly established using RVs determined by means of the first moment, bisector or double Gaussian techniques applied to the H$\alpha$ emission line \citep[e.g.][and references therein]{Naz22}. The first order moment is especially sensitive to the morphology of the whole profile. The other methods give more weight to the line wings. The mirror method compares the blue wing to the red wing mirrored about a centroid velocity. The best estimate of the centroid velocity is obtained when both wings overlap. The double Gaussian method consists in correlating the profiles with two Gaussians of same width but opposite signs and opposite centre velocities. The correlation is computed for different shifts of the function, and the best estimate of the RV is found when the correlation is equal to zero. Owing to the very strong line profile variations, one must be cautious when applying these methods to HD\,45314. We focus on the results obtained with the double Gaussian method which should be less affected by the variability of the line profile. The resulting RVs fail to indicate coherent trends typical of orbital motion. The RVs seem much affected by the line profile variations, with larger changes recorded before 2023. This is especially marked during the oscillation phase, when RVs vary between $-40$ and $+60$\,km\,s$^{-1}$, while orbital motion of OBe stars typically has $K = 5$ -- $10$\,km\,s$^{-1}$ amplitude. Since 2023, with the disc slowly rebuilding, the RV changes appeared smaller, with peak-to-peak variations less than $25$\,km\,s$^{-1}$ but again no obvious signature of orbital motion. Nevertheless, we performed a Fourier analysis of these RVs, and the strongest peaks correspond to periods of $337.8 \pm 2.1$\,d (high state), $333.3 \pm 20.6$\,d (shell phase), $454.5 \pm 12.8$\,d (oscillation phase) and $416.7 \pm 15.2$\,d (disc rebuilding phase). The disagreement between these "periods", combined to their poor significance and the unconvincing variations in the folded RVs, argue against the detection of orbital motion.

Whilst HD\,45314 presents some absorption lines in its spectrum (e.g.\ He\,{\sc ii} $\lambda$\,4686), they are also contaminated to some extent by the wings of neighbouring emission lines and could also be affected by intrinsic variations. Neither the RVs of the He\,{\sc ii} $\lambda$\,4686 line, nor those obtained via cross-correlation, display a coherent behaviour. These RVs vary over a rather wide range, and a Fourier analysis yields a periodogram essentially consistent with white noise, lacking any outstanding peak. Therefore, at this stage, neither the emission lines nor the absorption lines reveal a binary signature.  

\section{Short-term variations \label{short}}
        This section addresses photometric variations on timescales between a few hours and several days. We also describe how the amplitudes of these modulations vary over time.
        
\begin{figure}[h]
    \begin{center}
          \resizebox{8.5cm}{!}{\includegraphics{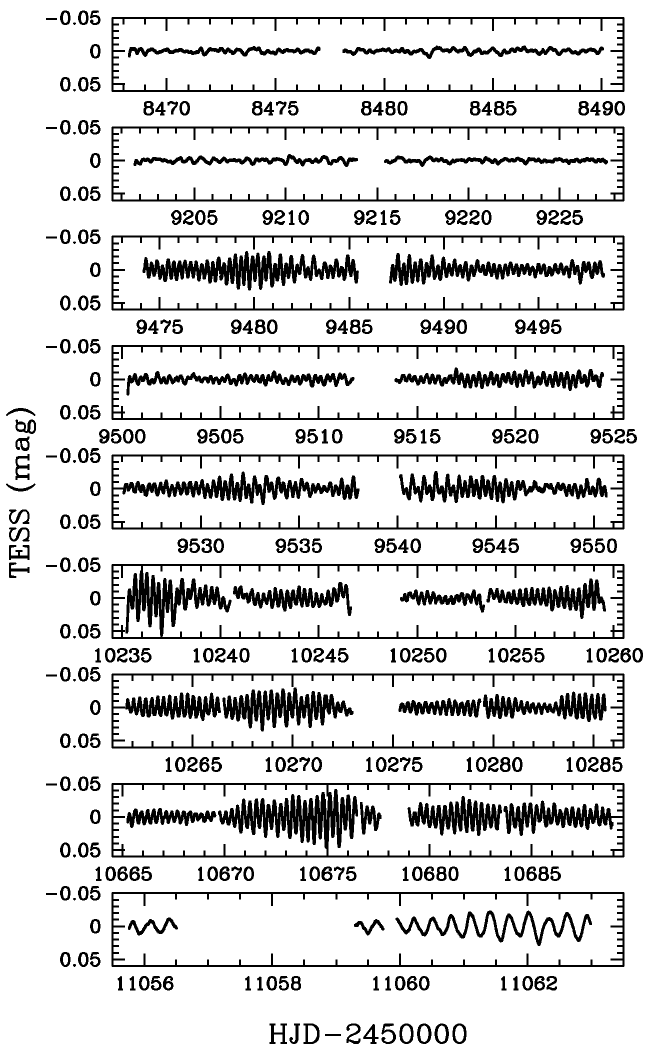}}
          \caption{Detrended {\it TESS} light curves of HD\,45314. From top to bottom, the panels correspond to Sectors\,6, 33, 43, 44, 45, 71, 72, 87 and 1751. \label{TESSlc}}
    \end{center}
\end{figure}

The {\it TESS} data provide a unique dataset to probe the short-term photometric variability and its evolution with disc state. In addition to long-term changes between sectors, changes on intermediate timescales, during one sector, can also be spotted in the light curves (see Figs.\,\ref{TESSvsKWS} and \ref{detrending} in Appendix\,\ref{trendsTESS}). Since we are interested in the short-term behaviour of HD\,45314, such medium- to long-term trends need to be removed. To this aim, we subtracted the average magnitudes from light curves of Sectors\,6, 33 and 1751: this was sufficient as they were quite stable over the duration of the month-long observation. For the other, more variable, sectors, a moving average with size $\pm1$\,d was computed and then subtracted from the light curves (see Fig.\,\ref{detrending}). Figure\,\ref{TESSlc} displays the detrended {\it TESS} light curves.

\begin{figure*}[h]
  \begin{minipage}{6cm}
    \resizebox{6cm}{!}{\includegraphics{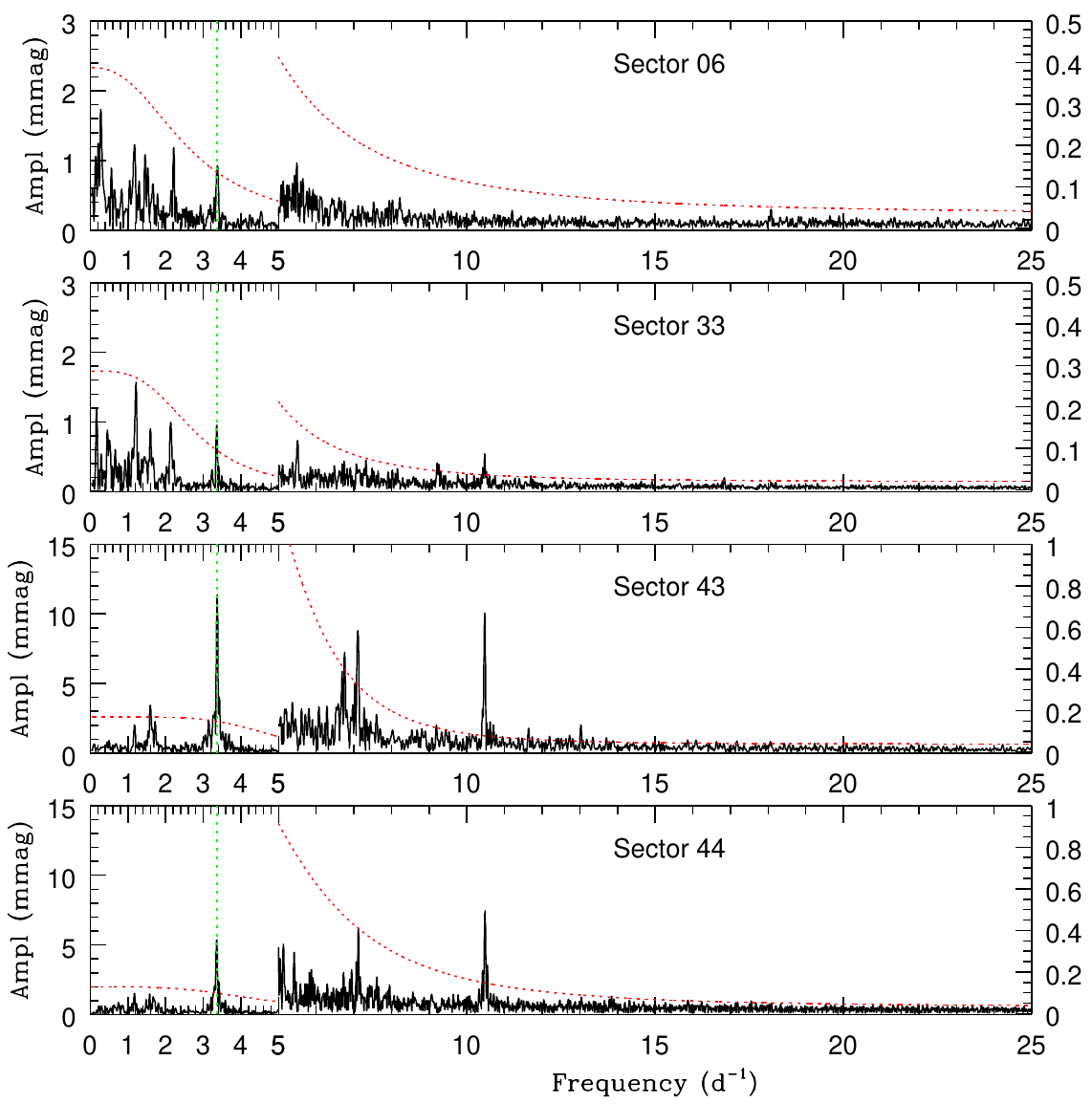}}
  \end{minipage}
  \hfill
  \begin{minipage}{6cm}
    \resizebox{6cm}{!}{\includegraphics{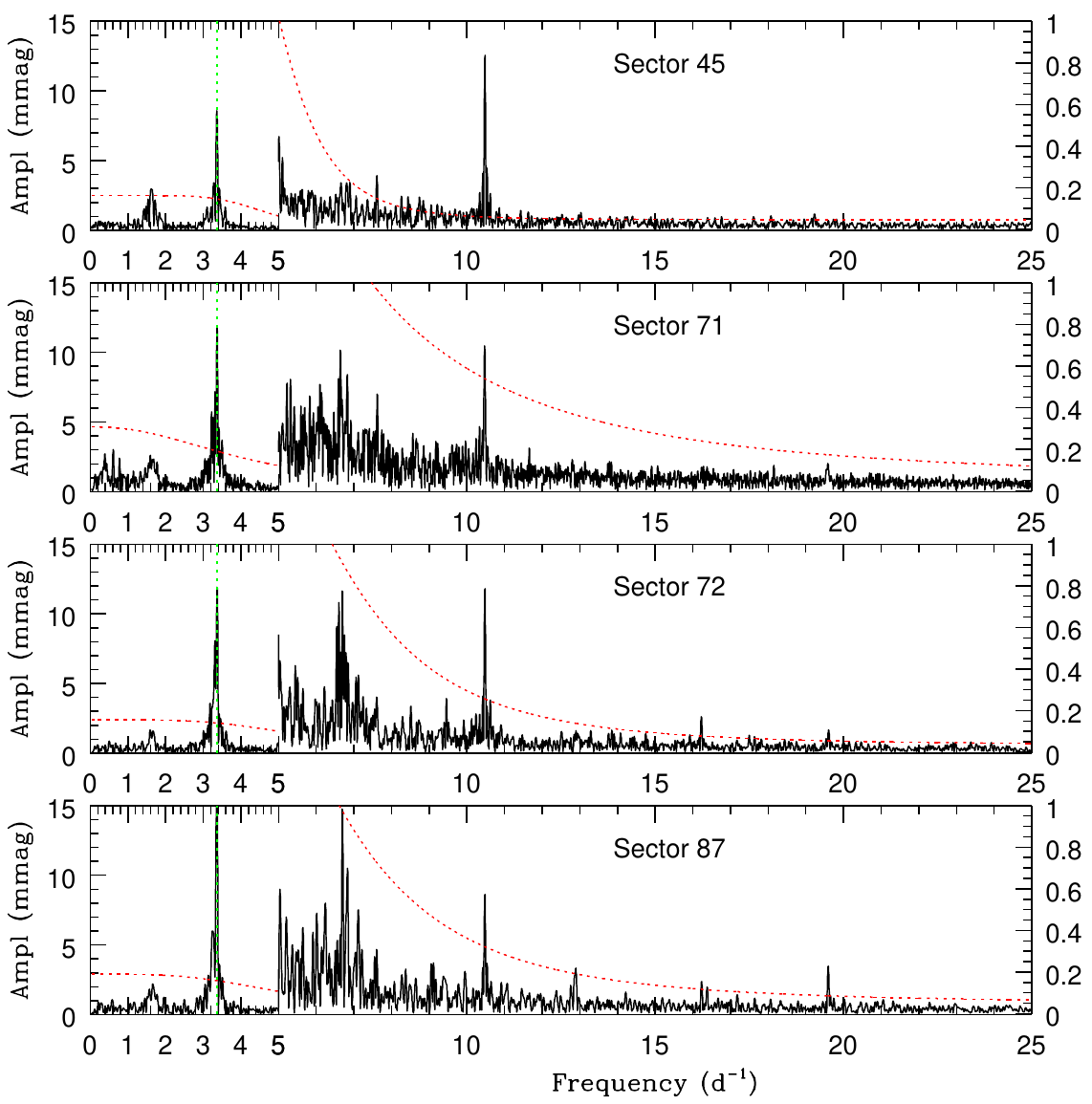}}
  \end{minipage}
  \hfill
  \begin{minipage}{6cm}
    \resizebox{6cm}{!}{\includegraphics{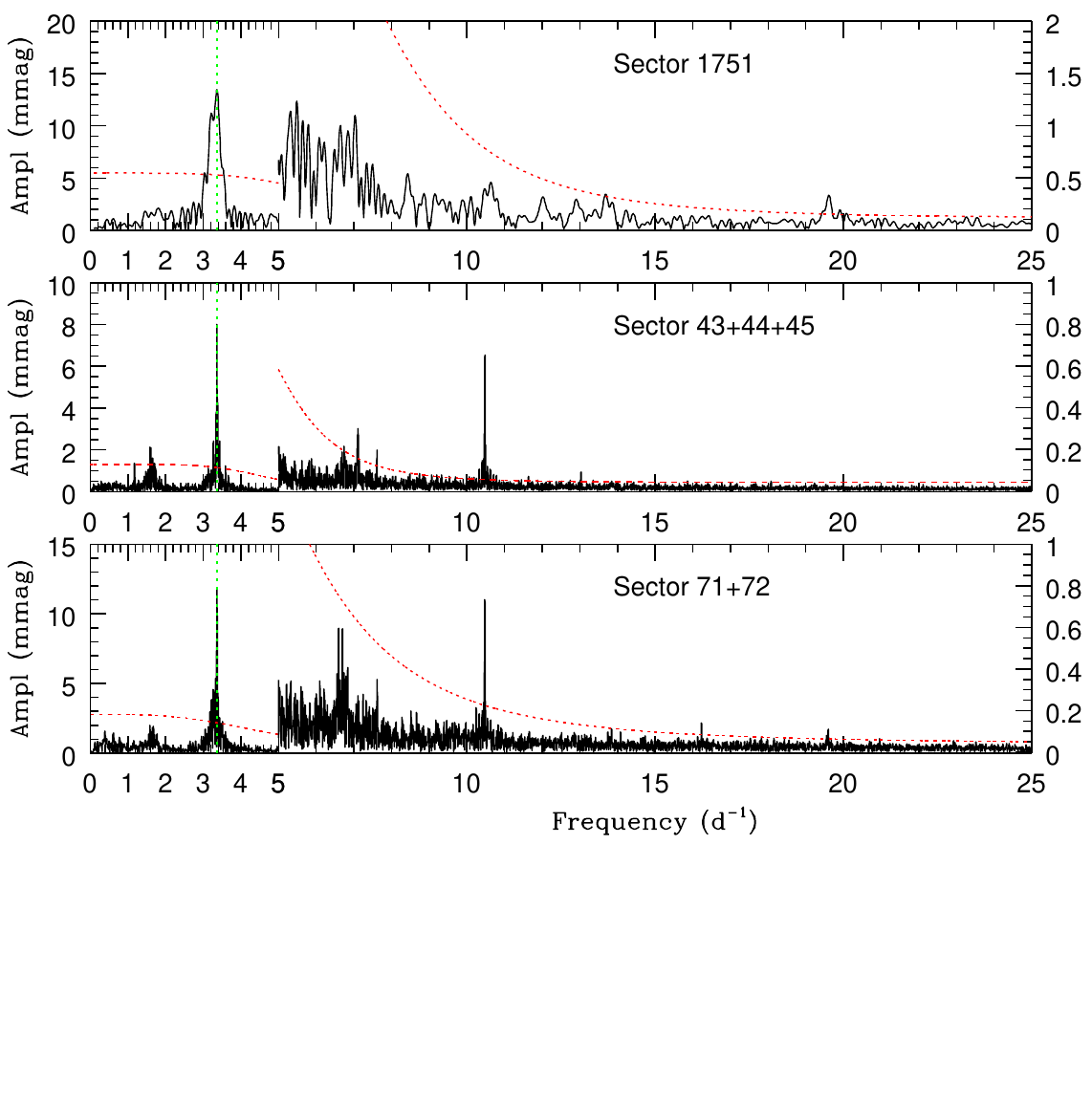}}
  \end{minipage}
  \caption{Fourier periodograms of the {\it TESS} light curves. To emphasize the visibility of features at higher frequencies, the vertical scale of the plots changes at a frequency of 5\,d$^{-1}$, switching from the left vertical axis to the right one. The red dotted curve provides the significance threshold defined as five times the red + white noise model. The dotted green line indicates $\nu_1$.\label{Fourier_TESS}}
\end{figure*}

The data of Sectors\,6 and 33 were previously analysed in \citet{Naz20} and \citet{Lab21}. They display variability of relatively low amplitude. As shown in Figs.\,\ref{historic} and \ref{TESSvsKWS}, these observations were collected during the oscillation phase at a time when the star was brightening at a very slow rate (Sector\,6) or was near maximum brightness (Sector\,33). The situation significantly changed in the subsequent observations where the overall brightness was lower and the photometry exhibited significantly larger short-term variations. Sectors\,43 -- 45 covered the oscillation that marked the transition to the decline phase, whilst Sectors\,71 -- 72, 87 and 1751 sampled the minimum and slow recovery phase. Considering the longest continuous interval covered by {\it TESS}, that is Sectors\,43 -- 45, we note that the brightness at first steeply increased from m$_{\rm TESS} \simeq 6.9$ to $6.7$ throughout Sector\,43. During Sector\,44, the magnitude remained quite stable near 6.65 and finally evolved to $m_{\rm TESS} \simeq 6.75$ over Sector\,45. Quite interestingly, the data from Sectors\,43 and 45 exhibit the largest short-term variations, whilst data from Sector\,44 display short-term variations of lower amplitude. A similar trend of more prominent short-term variations when the star is globally fainter is observed during Sectors\,71 \& 72.

Fourier frequency spectra were computed for all light curves up to the Nyquist frequency but were in fact empty at frequencies above 20\,d$^{-1}$. All periodograms clearly showed the signature of prominent low-frequency stochastic variations, so-called red noise. We adjusted a model for the red + white noise using the formalism of \citet{Sta02} and cutting out individual peaks of high amplitude (typically above 2\,mmag). The adjusted expression was
\begin{equation}
  A(\nu)=C+\frac{A_0}{1+(2\pi \tau \nu)^\gamma},
\end{equation} 
with $C$ the white-noise level, $A_0$ the red-noise level at null frequency, $\tau$ the mean lifetime of the structures producing the red noise, and $\gamma$ the slope of the linear decrease between red and white noise. Since some light curves were detrended, the formal values of the noise parameters may be affected, and will not be discussed further. The noise model simply allows us to assess the significance level of individual peaks: they are deemed significant if their amplitude exceeds five times the level of the red + white noise model. 

\begin{table}
  \caption{Significant or recurrent signals detected in the Fourier periodograms of the {\it TESS} photometry.\label{TESSfreq}}
  \begin{tabular}{l l}
    \hline
    \multicolumn{1}{c}{$\nu$ (d$^{-1}$)} & \multicolumn{1}{c}{Remarks} \\
    \hline
    $\sim 1.6$ & faint FG (Sectors\,43, 45, 71, 72, 87) \\
    3.364 -- 3.380 & $\nu_1$ (all Sectors) \\
    5.596 & (only Sector\,71) \\
    6.600 -- 6.756 & (Sectors\,43, 71, 72, 87), $2\,\nu_1 = 6.738$\,d$^{-1}$ \\
    7.108 -- 7.124 & (Sectors\,43, 44) \\
    7.616 -- 7.620 & (Sectors\,45, 71 -- 72) \\    
    9.220 & (only Sector\,33) \\
    10.476 -- 10.488 & (Sectors\,33, 43, 44, 45, 71, 72, 87) \\
    12.896 & (only Sector\,87) \\
    16.232 -- 16.236 & (Sectors\,72, 87) \\
    16.840 & $\sim 5\,\nu_1$ (Sector\,33) \\
    18.085 & (Sectors\,6, 43 -- 45) \\   
    19.600 -- 19.620 & (Sectors\,72, 87, 1751) \\
    \hline
  \end{tabular}
  \tablefoot{Typical errors on the frequencies are $0.004$\,d$^{-1}$.}
\end{table}

The most striking individual feature is a stable signal at a frequency near $\nu_1 = 3.369$\,d$^{-1}$. The amplitude of this signal varies with time between 0.9\,mmag (Sector\,6) and 16\,mmag (Sector\,87), although it is always visible. No unambiguous harmonics are visible, except perhaps for the fifth harmonic during Sector\,33. We examined more closely the peak profile in the longest dataset (combination of Sectors\,43 -- 45): smaller peaks are flanking the main one, but they correspond to aliases, as revealed by the spectral window. Therefore, this signal appears isolated. The second most stable peak is found at 10.480\,d$^{-1}$. It is detected in all but the first and last Sectors. With an amplitude of 0.09 -- 0.8\,mmag, it is the second strongest signal in the frequency spectra, and it is not an harmonic of $\nu_1$.

Whilst the frequency group (FG) near 1.6\,d$^{-1}$ has an amplitude that remains at the limit of significance, its recurrent presence motivated us to include it in Table\,\ref{TESSfreq}. Such low-frequency groups are common in Be stars \citep{Lab21,Lab22,Naz20} but they are here much weaker than the isolated signals. Additional high-frequency features are repeatedly present (and often significant) in several Sectors, see Table\,\ref{TESSfreq}.

\begin{figure}
  \begin{center}
    \resizebox{8.7cm}{!}{\includegraphics{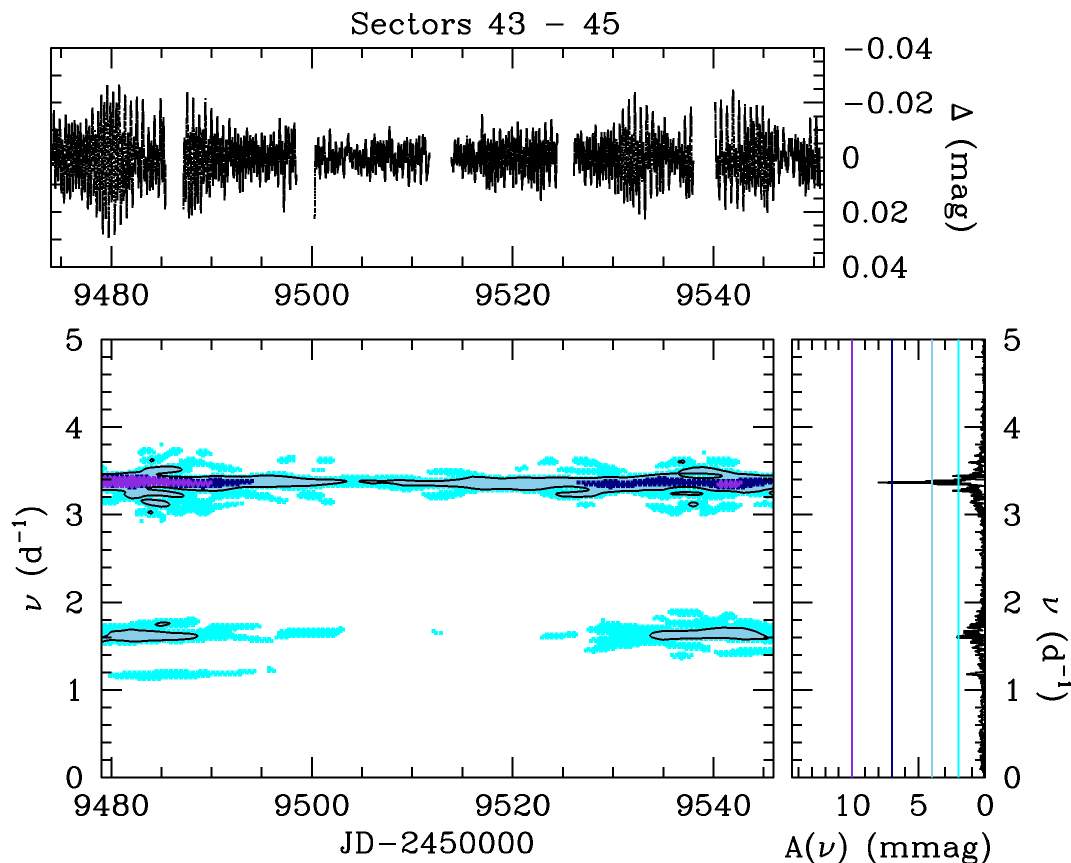}}
  \end{center}
  \caption{Time–frequency diagram of the {\it TESS} photometry of HD\,45314 from Sectors\,43 -- 45. The detrended light curve is shown in the top panel. The colour scale diagram illustrates the evolution of the Fourier periodogram as a function of the date corresponding to the middle of a 10\,d sliding window. The violet, dark blue, light blue and cyan colours respectively represent areas with amplitudes $\geq 10$, $\geq 7$, $\geq 4$ and $\geq 2$\,mmag. The bottom right panel provides the periodogram evaluated with the entire dataset of Sectors\,43 -- 45. The coloured lines correspond to the scale used in the bottom left panel.\label{spevol7172}}
\end{figure}

For the combined Sectors\,43 -- 45 and 71 -- 72, we further built time-frequency diagrams by computing Fourier periodograms of the photometric data extracted over a sliding temporal window of 10\,d duration and shifted in steps of 1\,d. Figure\,\ref{spevol7172} illustrates the temporal evolution of the Fourier periodogram up to 5\,d$^{-1}$ during the combined Sectors\,43 -- 45. We can clearly see that the amplitude of the dominant mode changes on timescales of about a week. 

In Appendix\,\ref{appshort}, we analyse the spectra collected during our intensive TIGRE and Aur\'elie campaigns to search for short-term spectroscopic variations. Whilst the TIGRE and 2020 Aur\'elie campaigns do not overlap with any of the {\it TESS} observations, the 2021 Aur\'elie campaign (HJD\,2459493.6 -- 2459498.7) falls right in the second half of Sector\,43. Significant line profile variations were detected in the emission lines for each of these intensive spectroscopic campaigns. However, no significant periodic signals, such as expected from non-radial pulsations occurring in the photosphere, were found. The detected variations rather seem connected to longer or medium-term phenomena affecting the circumstellar disc. Furthermore, the analysis of the RVs collected during those campaigns revealed that the bulk of the long-term RV dispersion stems from measurement uncertainties and stochastic variations taking place on short timescales.    

\begin{figure}
  \begin{center}
    \resizebox{8.7cm}{!}{\includegraphics{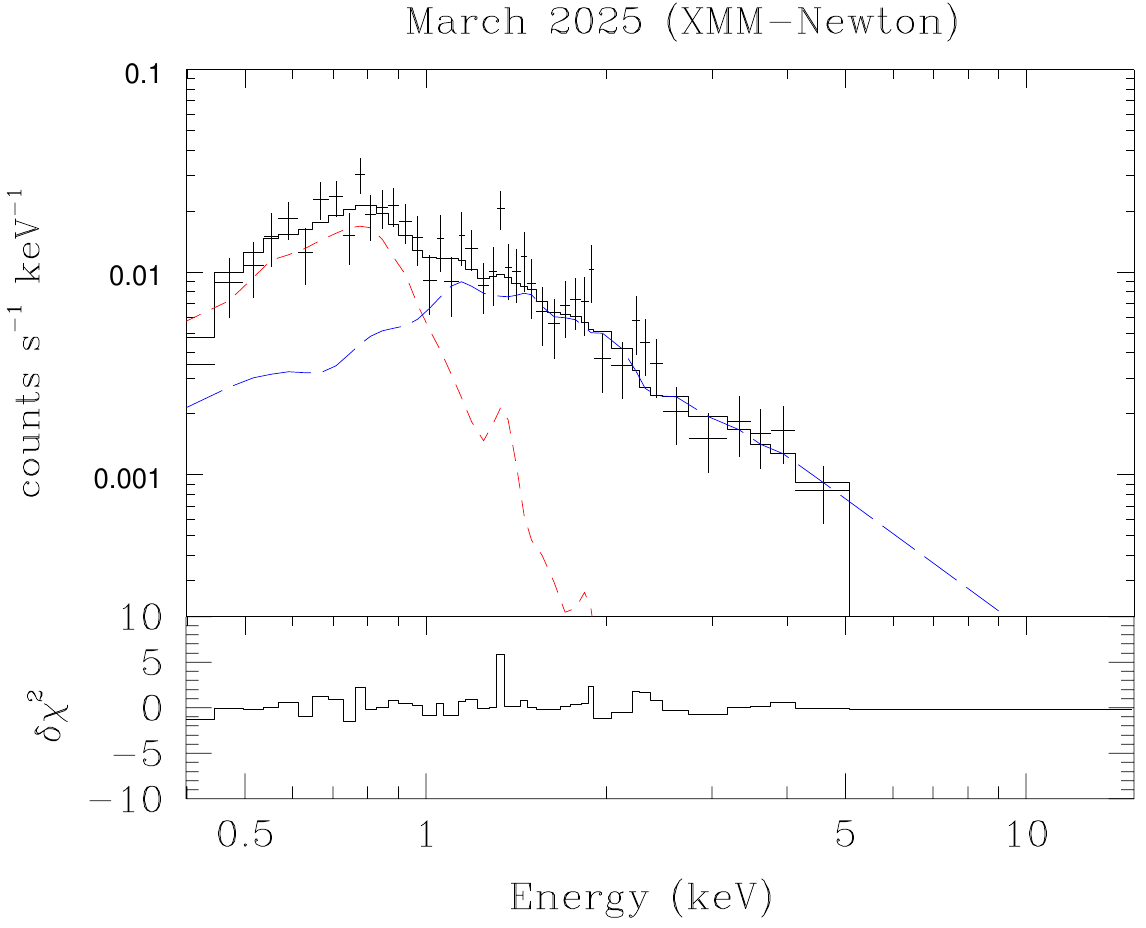}}
  \end{center}
  \caption{{\it XMM-Newton} EPIC-pn spectrum of HD\,45314 in March 2025. The black symbols with error bars stand for the observed spectrum whereas the black histogram displays our best-fit model from Table\,\ref{fitX}. The dashed red and blue curves stand respectively for the soft and the hard plasma component. The lower panel illustrates the contributions of individual energy bins to the overall $\chi^2$ with the sign corresponding to the sign of the data minus model.\label{specEPIC}}
\end{figure}

\begin{table*}
  \caption{Results of our analysis of the March 2025 (HJD\,2460745.669) EPIC spectra of HD\,45314.\label{fitX}}
  \begin{center}
  \begin{tabular}{c c c c c c c c c c c}
      \hline
      \multicolumn{11}{c}{{\tt tbabs*wind*apec(2T)}}\\
      \hline
      $\log{N_{\rm wind}}$ & $kT_h$ & Norm$_h$ & $kT_s$ & Norm$_s$ & $\chi^2_{\nu}$ & d.o.f. & $f_{\rm X}^{\rm soft}$ & $f_{\rm X}^{\rm med}$ & $f_{\rm X}^{\rm hard}$ & $f_{\rm X}^{\rm int}$\\
      \vspace*{-3mm}\\
      \cline{8-11}
      \vspace*{-3mm}\\
      (cm$^{-2}$) & (keV) & ($10^{-5}$\,cm$^{-5}$) & (keV) & ($10^{-5}$\,cm$^{-5}$) & & & \multicolumn{4}{c}{($10^{-14}$\,erg\,cm$^{-2}$\,s$^{-1}$)} \\
      \hline
      \vspace*{-3mm}\\
      $21.49^{+0.39}$ & $5.2^{+3.0}_{-1.3}$ & $7.4^{+1.3}_{-1.1}$ & $0.27^{+0.03}_{-0.03}$ & $6.6^{+15.8}_{-4.6}$ & 0.89 & 76 & $1.8 \pm 0.2$ & $2.1 \pm 0.2$ & $7.3 \pm 1.2$  & $13.5 \pm 1.2$\\  
      \vspace*{-3mm}\\       
        \hline
  \end{tabular}
  \end{center}
  \tablefoot{The fit includes a fixed neutral hydrogen column density of $1.9 \times 10^{21}$\,cm$^{-2}$. The normalisation parameters of the {\tt apec} models correspond to $\frac{\int n_e\,n_{\rm H}\,dV}{4\,\pi\,d^2}$, with $d$ the distance to the source (in cm), $n_e$ and $n_{\rm H}$ the electron and proton density of the emitting plasma. Columns 8 to 10 list the observed fluxes evaluated over the soft (0.5 -- 1.0\,keV), medium (1.0 - 2.0\,keV) and hard (2.0 - 10.0\,keV) energy band. The last column provides the flux in the 0.5 - 10\,keV band corrected for the absorption by the interstellar medium ($N_{\rm H} = 1.9 \times 10^{21}$\,cm$^{-2}$).}
\end{table*}
\section{X-ray emission \label{SectX}}
To characterise the X-ray emission of HD\,45314 and complement the previous data, we investigate here the recent {\it XMM-Newton} and {\it eROSITA} data. We analysed the latest {\it XMM-Newton} observation of HD\,45314 using version 12.11.1 of the {\tt xspec} software\footnote{For consistency, the older {\it XMM-Newton} and {\it Suzaku} data were also re-analysed with this version of the software. The results were identical to those of \citet{Rau18} within the error bars.} \citep{Arn96}. To ease comparison with previous studies, and because it provides a good adjustment of the spectra, we used an absorbed two temperature plasma model. Interstellar absorption, with a neutral hydrogen column density of $1.9 \times 10^{21}$\,cm$^{-2}$ was modelled with the T\"ubingen-Boulder model \citep[{\tt tbabs},][]{Wil00}. Additional circumstellar absorption by ionized stellar wind material was treated with the stellar wind absorption model ({\tt wind}) of \citet{Naz04}. Finally, the emission from optically thin thermal plasma was represented by means of the {\tt apec} model \citep{APEC}. The results are quoted in Table\,\ref{fitX}. Fitting the observed spectrum required two thermal plasma components: one for the softer part of the spectrum ($kT_s \simeq 0.27$\,keV), and another one for the harder part ($kT_h \simeq 5.2$\,keV, see Fig.\,\ref{specEPIC}). The March 2025 spectrum and its best-fit parameters are remarkably similar to the March 2016 data \citep{Rau18}. The small $\sim 9$\% decrease of the observed flux between March 2016 and March 2025 remains within the uncertainties on the  fluxes.  

In the {\it eROSITA} All Sky Survey (eRASS) source catalogue, HD\,45314 has a count rate of $0.103 \pm 0.018$\,ct\,s$^{-1}$ in the 0.2 -- 2.3\,keV band for the combined data from the four half-year surveys. Despite rather large error-bars on data from individual surveys, the source appears variable, with the count rates of individual surveys ranging between $0.025 \pm 0.021$\,ct\,s$^{-1}$ in spring 2021 and $0.199 \pm 0.046$\,ct\,s$^{-1}$ six months later (see also Fig.\,\ref{historic}). Using an absorbed 2-T plasma model to fit the combined {\it eROSITA} spectrum of HD\,45314 resulted in best-fit temperatures and circumstellar columns that overlap within the errors with those in Table\,\ref{fitX}. The {\it eROSITA} spectrum indicates a somewhat higher flux than the March 2025 {\it XMM-Newton} data, especially in the 1.0 -- 2.0\,keV band. The global (0.5 -- 10\,keV) flux corrected for interstellar absorption inferred from the {\it eROSITA} spectrum amounts to $17.4^{+19.6}_{-11.2}\,10^{-14}$\,erg\,cm$^{-2}$\,s$^{-1}$, that is 30\% higher than in the March 2025 {\it XMM-Newton} data. 

In Appendix\,\ref{app3}, we take advantage of the blue spectrum observed at the minimum emission state to revise the spectral classification to O9.7-B0\,V, leading to a bolometric flux of $9.7 \times 10^{-7}$\,erg\,cm$^{-2}$\,s$^{-1}$. This allows us to re-derive the $\log{L_{\rm X}/L_{\rm bol}}$ values. We obtain values of $-5.93$ for the high-state {\it XMM-Newton} spectrum and $-6.17$ for the {\it Suzaku} shell-phase observation, whilst the low-state {\it XMM-Newton} observations yield $-6.89$.   

Recently, \citet{Web26} reported a recurrent periodicity of about 5.4\,ks in the previous two {\it XMM-Newton} datasets (April 2012 and March 2016). We used the March 2025 observation to search for variations in the X-ray emission at frequencies below 1\,mHz. Beyond Fourier spectra, computed on light curves extracted with bins of 10\,s or 100\,s, we also applied the same epoch-folding techniques as used by \citet{Web26}. For the March 2025 {\it XMM-Newton} observation, these techniques yield peaks around 3.8 and 6.8\,ks. It thus seems that, while significant periodicities can be identified in each exposure, they are not stable. Indeed, the frequency spectrum (whatever the technique used to compute it) varies from one exposure to the other.

\section{Discussion}
Over the past three decades, the electromagnetic signatures of the circumstellar envelope of HD\,45314 changed dramatically from an optically bright, emission-line-dominated, spectrum in the early years of this century to a much dimmer, essentially emission-free, spectrum in 2024. This is probably not the first time that the star lost most of its disc. Indeed, according to \citet{Cop63}, double-peaked H$\beta$ and H$\gamma$ emissions were seen in 1939, as well as between 1954 and 1961, but had disappeared in 1962. However, the sampling of this earlier event was very scarce and no quantitative measurements (EWs, peak separations, etc.) are available. Thanks to our intensive monitoring campaign, we have a much more detailed view of the most recent event.

\subsection{Long- and medium-term variations}
If the $\gamma$\,Cas properties of HD\,45314 stem from an accreting WD companion, this implies that the star is a binary system and we must interpret its behaviour in the context of binarity with a presumably low $q = \frac{m_{\rm companion}}{m_{\rm Be}}$ mass ratio. We thus start by confronting our observations to the expectations of smooth particle hydrodynamics (SPH) simulations for Be binary systems.

\citet{Pan18} used SPH calculations to show that in Be binary systems the tidal action of a coplanar companion can lead to the formation of a two-armed density enhancement in the Be disc. Such a spiral structure would manifest itself via V/R modulations occurring with a period half the orbital period, $P_{\rm orb}$. An alternative mechanism to produce cyclic V/R variations are one-armed density waves due to disc oscillations \citep{Oka91} that do not require the action of a companion to take place. Interpreting the V/R cycles of HD\,45314 in the context of the \citet{Pan18} model would imply an orbital period of 2380\,d. This is much longer than the timescale of oscillations in $m_V$, which could themselves be multiples of $P_{\rm orb}$ (see below). HD\,45314 would thus be similar to $\zeta$\,Tau, another $\gamma$\,Cas star, where the duration of the observed V/R cycle ($\sim 1400$\,d) has no obvious connection to the orbital period \citep[133\,d,][]{Ste09}. The most likely explanation of the V/R cycles would thus be one-armed density waves due to disc oscillations \citep{Oka91} bearing no connection to binarity.

Previously, \citet{Rau15} and \citet{Rau18} attempted to adjust observed H$\alpha$ line profiles using a simple disc model. During the high state, the inferred apparent disc inclination was found to be near $40^{\circ}$ -- $45^{\circ}$, but increased to $80^{\circ}$ during the shell state and dropped to $\sim 15^{\circ}$ at times when the emission profile was very narrow. This surprising change in disc inclination might reflect a precession of the disc. \citet{Suf22} presented 3-D SPH simulations of the formation and dissipation of a Be disc in an equal mass ($q = 1$) Be binary with an orbital period of either 30 or 300\,d. The plane of the orbit was taken to be misaligned by $20^{\circ}$, $40^{\circ}$ or $60^{\circ}$ with respect to the equatorial plane of the Be star. \citet{Suf25} extended this work to $q$ values of 0.5 and 0.1. \citet{Suf22} found that during disc growth, for which they assumed mass injection in the equatorial plane, the companion tilts the disc away from the equatorial plane. This can lead to transitions between single-peaked and double-peaked line profiles \citep{Suf23}. Once mass injection stops in the SPH calculations, the disc precesses about the orbital axis on a precession period between 20 and 50 times $P_{\rm orb}$. As a result, the orientation of the disc with respect to the observer oscillates, resulting in oscillations of the observables at half the precession period \citep{Suf23}. Depending on the exact configuration, EW(H$\alpha$) and $m_V$ are either correlated or anticorrelated \citep[see Fig.\,14 of][]{Suf23}. 
Eccentric discs, undergoing Kozai-Lidov cycles, were also found in the SPH simulations. They go along with strong variations in the H$\alpha$ emission line profile \citep{Suf22,Suf25}. For a $40^{\circ}$ misalignment, \citet{Suf22} observed disc tearing in their simulations, where the outer disc detaches from the inner part. Depending on the orientation of the observer's sightline, this can also lead to anticorrelations between the EW(H$\alpha$) and $m_V$ \citep{Suf24}. Whether the disc undergoes tilting followed by precession or tearing, the behaviour of the observable quantities strongly depends on the orientation of the observer's sightline \citep{Suf23,Suf24}.   

Many of the properties of HD\,45314 are reminiscent of the predictions of SPH calculations for misaligned binary systems. Indeed, our analysis revealed that the variations in emission line EWs and optical brightness are correlated on timescales of decades but exhibit anticorrelations on timescales of a few years, except during the disc dissipation and low state. Part of the anticorrelation between the EWs and the optical brightness during the oscillation phase probably stems from the dilution of the circumstellar emission by a variable continuum. Assuming the emission to remain constant, the changes in the measured EW at time $t$ can be expressed as
\begin{equation}
  EW(t) = \left(1 - \frac{{\cal F}_c(t_0)}{{\cal F}_c(t)}\right)\,EW_{\rm abs}+\frac{{\cal F}_c(t_0)}{{\cal F}_c(t)}EW(t_0)
\end{equation}
where $EW(t_0)$ is the measured EW at time $t_0$, $EW_{\rm abs}$ is the EW of the photospheric absorption that we assume to remain constant, and ${\cal F}_c$ stands for the flux in the continuum. Using the contemporaneous $m_V$ and EW measurements, we estimated that, during the oscillation phase, about 80\% of the variations of EW(H$\beta$) simply reflect the dilution effect. For EW(H$\alpha$) and EW(He\,{\sc i} $\lambda$\,5876), we find significantly lower fractions of $\sim 40$\% and $\sim 30$\%, respectively. For those lines, additional factors must contribute to the observed anticorrelation.

It is thus tempting to interpret the oscillations of HD\,45314 as resulting from disc precession similar to the predictions of \citet{Suf22,Suf25}. Optical brightness would reach a maximum for the disc orientation closest to pole-on, whereas it would be minimum for the most equator-on configuration. Disc precession could have a larger impact on $m_V$ than on EW(H$\alpha$) as part of the inner disc could be occulted by the star at times, whilst the outer disc would be less affected by such occultations. In this case, the period of disc precession would be close to 460\,d. If the precession period corresponds to $\gtrsim 20\,P_{\rm orb}$, this would imply a relatively short $P_{\rm orb}$ of a few weeks\footnote{Known orbital solutions of $\gamma$\,Cas stars indicate orbital periods between about one month and half a year \citep{Naz22}.}. In such a model, variations in the disc orientation with respect to the observer go along with changes in the H$\alpha$ line morphology. If the outer radius of the H$\alpha$ emitting region remains constant, then changes in the disc orientation should manifest themselves as variations in the velocity separation between the emission peaks. Still, Fig.\,\ref{oscillations} demonstrates that the peak separation does not display a minimum when the star is brightest, that is when the disc is seen nearly pole-on, as one would expect. This casts some doubt on the precession interpretation. 

As an alternative, \citet{Lab17} proposed that low-frequency pulsations of the central star or coupling of several pulsation modes could explain photometric variability on timescales between a few days and about 200\,days. Last, but not least, tidally excited oscillations \citep[TEOs,][]{Kol23} could play a role in the photometric variability of the Be star and the oscillations of the disc. This mechanism, however, requires an eccentric orbit whilst existing orbital solutions of Be stars (including $\gamma$\,Cas analogs) all have a negligible eccentricity \citep{Naz22}.

Finally, we note that the optical brightness increased during the shell event (see Fig.\,\ref{historic}). This is surprising given that H$\alpha$ line profiles  were consistent with an inclination near $80^{\circ}$. Under these circumstances, one would rather expect the brightness to decrease due to obscuration of the photosphere by the Be disc. This behaviour hints at a warped disk configuration, where obscuration by material in front of the star is compensated by reflection and reprocessing of stellar radiation by the rear part of the disk that would be seen under a lower viewing angle. 

\subsection{Pulsation properties}
Our analysis of {\it TESS} data of HD\,45314  revealed a number of potential $\beta$\,Cep-type pulsation frequencies. Given the star's location in the $\beta$\,Cep instability strip \citep[see Fig.\,8 of][]{Naz20}, the presence of multi-periodic pulsations in the photometry of HD\,45314 is not a surprise. Comparing with other rapidly rotating pulsating late O-type stars (e.g.\ $\zeta$\,Oph, \citealt{How14}; HD\,93521, \citealt{Rau21b}; HD\,256035, \citealt{Buy15}), HD\,45314 and $\zeta$\,Oph are those with the largest amplitudes of photometric variations.  

The amplitude of HD\,45314's main mode, at $\nu_1 = 3.369$\,d$^{-1}$, changed between 0.9 and 16\,mmag on timescales of weaks or months. This situation is reminiscent of the strongest signals in spaceborne photometry of the O9.5 star $\zeta$\,Oph. For this star, \citet{How14} found variations in pulsation amplitudes between $\leq 2$\,mmag and $> 20$\,mmag on timescales of hundreds of days. Such variations in amplitude (and frequency) could result from non-linear resonant coupling between pulsation modes of multiperiodic stars \citep[e.g.][]{Pig08}.

\citet{Naz20b} noted variations in the amplitude of the dominant pulsation modes of the $\gamma$ Cas star $\pi$\,Aqr as a function of disc activity. It is thus interesting to examine the overall level of HD\,45314's photometric variability as a function of the state of its circumstellar disc. Indeed, non-radial pulsations are often considered a necessary ingredient of the Be disc formation process \citep[][and references therein]{Riv13}. Most of HD\,45314's {\it TESS} data sample the oscillations or decline phase. For those epochs, we note that the amplitude of short-term modulations is largest at times when the medium- and long-term brightness variations display the steepest gradient (Sectors\,43, 45 \& 87). Sectors\,71, 72, 87 and 1751 sample the early phases of the current disc build-up phase: during Sector\,87, we observe the largest amplitude of the $\nu_1$ mode in any of our datasets. Conversely, very low amplitudes of short-term variations are found at times when the long-term variations flatten having reached a brightness maximum (Sectors\,6 and 33). 

Our intensive spectroscopic monitoring campaigns did not reveal obvious spectroscopic counterparts of these photometric variations. This situation contrasts with the case of $\pi$\,Aqr where rapid ($\sim 11.8$\,d$^{-1}$) $\beta$\,Cep-type variations were seen both in photometry and spectroscopy \citep{Naz20b}. 

\subsection{The X-ray emission \label{Xrayanalysis}} 
When first observed with {\it XMM-Newton} in April 2012, HD\,45314 displayed a clear $\gamma$\,Cas-like SED \citep{Rau13}. This was consistent with a previous detection of the star with {\it EXOSAT} in October 1984 (see Appendix\,\ref{appXother}). During the shell episode, in October 2014, {\it Suzaku} recorded a spectrum that still indicated a $\gamma$\,Cas-like SED, but with an X-ray flux half of that observed two and a half years before. The second {\it XMM-Newton} observation of HD\,45314, collected during an episode of substantial fading of the H$\alpha$ emission in March 2016, revealed a major softening and dimming of the X-ray spectrum \citep{Rau18}. All subsequent X-ray observations ({\it eROSITA}, {\it XMM-Newton} discussed in Sect.\,\ref{SectX} and {\it Swift} discussed in Appendix\,\ref{appXother}) showed the star to be in this low state. Figure\,\ref{historic} shows that the transition of the X-ray emission happened well before the strongest dissipation of the decretion disc as diagnosed from the optical brightness and H$\alpha$ emission strength. If we interpret this in the context of an accreting WD, it implies that accretion ceased, or at least decreased tremendously, whilst most of the Be disc was still present. If the putative WD accretes material via a kind of disc Roche lobe overflow \citep{Ras25}, then this situation probably implies that the density of the outer regions of the Be disc dropped below a critical value.  

The behaviour of the X-ray emission of HD\,45314 as a function of the properties of the Be disc contrasts with that of some other $\gamma$\,Cas stars. \citet{Naz19b} investigated the X-ray emission of $\pi$\,Aqr at times when the star was in a low nearly disc-free state and when it was in a strong emission state. Likewise, \citet{Naz22b} compared the X-ray emission of HD\,119682 in 2019 -- 2021, as its H$\alpha$ emission was fading and completely disappearing at some times, to earlier X-ray observations. $\pi$\,Aqr displayed a roughly 50\% lower X-ray flux when the optical disc diagnostics were low. This reduction is much smaller than for HD\,45314. Also, in the case of $\pi$\,Aqr, V767\,Cen, and HD\,119682, the hardness of the X-ray emission only slightly decreased or remained constant when H$\alpha$ emission was low. This is at odds with the considerable softening observed in the case of HD\,45314. These differences might indicate that HD\,45314 has a wider orbit than $\pi$\,Aqr ($P_{\rm orb} = 84$\,d) and HD\,119682 ($P_{\rm orb} = 59$\,d, Naz\'e et al.\ 2026b, submitted), thereby rendering disc Roche lobe overflow more difficult. Alternatively, the earlier spectral type, and thus stronger radiation field, of HD\,45314 could lead to a faster disc ablation \citep{Kee16}, thereby affecting the mass transfer to the companion.  

However, though the overall X-ray brightness and hardness have dropped tremendously, there still remains some residual hard emission, even now that the disc has been dissipating for several years. Our most recent {\it XMM-Newton} observation displays a plasma component with a rather high $kT_2 \simeq 5.2$\,keV and a luminosity of $9 \times 10^{30}$\,erg\,s$^{-1}$, which was also present in the March 2016 observation. Such a temperature is difficult to explain via typical intrinsic X-ray emission processes of massive stars, such as wind embedded shocks. In the context of an accreting magnetic WD companion \citep[see][and references therein]{Naz26}, the X-ray SED of the post-shock plasma is to first order ruled by the pre-shock velocity of the gas flowing along the WD's magnetic field lines. This quantity depends on the ratio between the WD's mass and radius, but also on the magnetospheric radius and on the height of the shock above the WD surface, $H_{\rm sh}$ \citep{Sul25}. The decrease in X-ray luminosity seen in HD\,45314 implies a reduction of the WD accretion rate, which goes along with an increase of $H_{\rm sh}$ \citep{Sul25}. According to equation (8) of \citet{Sul25}, a taller accretion column implies a lower post-shock plasma temperature. The taller column also enhances cyclotron cooling, thus further reducing the plasma temperature. In this picture, the $\simeq 5.2$\,keV emission could arise from low-level residual accretion possibly fed by a slow wind via a Bondi-Hoyle-Lyttleton process. The wind involved in this process could either be the stellar wind of the Oe star or a disc wind arising from the disc ablation by the star's radiation field \citep{Kee16}.

\section{Conclusions}
        In this work, we have analysed a large set of spectroscopic, photometric and X-ray data of the (currently) dormant $\gamma$\,Cas star HD\,45314. Over the past three decades, HD\,45314 underwent strong changes in its optical and X-ray emission, which likely reflect the dissipation of the decretion disc. The optical disc diagnostics went through different phases, including an oscillation state where optical brightness and line strengths were modulated on a 230\,d period. The emission line profiles displayed V/R variations on a longer, less well-defined, timescale of 1190\,d. The long-term optical variations bear resemblance with the predictions of SPH calculations for a disc undergoing tidal interactions with a companion. However, a more detailed comparison requires knowledge of the orbital properties which are currently lacking because no clear signature of orbital motion could be found.

Unlike some other $\gamma$\,Cas stars, where the X-ray emission remained nearly constant although the disc suffered major changes, HD\,45314's X-ray properties changed dramatically once the disc dissipation began. The X-ray emission changed from a hard and bright state to a much dimmer and softer state. This transition happened well before the bulk of the disc had dissipated. Assuming the hard X-ray emission to arise from accretion onto a putative WD companion, these different behaviours might hint at a wider orbital separation. As an alternative, the early stellar spectral type could influence the disc dissipation process, leading to a faster and more extreme reaction of the accretion flow. Despite the strong changes, the X-ray SED still contains a plasma component at a temperature near 5\,keV. This hot component hints at residual accretion, but at a significantly lower accretion rate than during the high state.

\begin{acknowledgements}
We dedicate this article to the memory of our colleague Jan Robrade who left us prematurely in July 2026. This work used spectra taken with the TIGRE telescope at La Luz Observatory (Guanajuato, Mexico). TIGRE was funded and operated by the universities of Hamburg, Guanajuato, and Li\`ege. We further used data collected by the {\it TESS} mission, publicly available from the Mikulski Archive for Space Telescopes. Funding for {\it TESS} is provided by NASA’s Science Mission directorate. This research also relied on data from {\it eROSITA}, the soft X-ray instrument on board {\it SRG}, and data supplied by the UK {\it Swift} Science Data Centre at the University of Leicester. ADS and CDS were used during this research. YN acknowledges support from the Fonds National de la Recherche Scientifique (Belgium). 
\end{acknowledgements}  

\bibliographystyle{aa}
  \bibliography{mybiblioAandA}
  \begin{appendix}
    \nolinenumbers
    \section{Further information on amateur spectroscopic equipment \label{Bess}}
    Table\,\ref{ProAm} summarizes the information about the telescopes and spectrographs used by the amateur astronomers who contributed to the HD\,45314 observing campaign. 
    \begin{table}[h!]
      \caption{Equipment used by amateur astronomers.\label{ProAm}}
        \resizebox{9cm}{!}{
        \begin{tabular}{l c c c c r}
        \hline
        Observer & Telescope & Spectrograph & R & Domain & N \\
        \hline
        E.\ Bryssinck & Celestron C11 0.28\,m & LHiResIII & 15000 & H$\alpha$ & 110 \\
        S.\ Charbonnel & Newton 0.5\,m & Eshel \#112 & 11000 & echelle & 1 \\
        2SPOT & RC12 0.30\,m & eShel & 11000 & echelle & 6 \\
        A.\ de Bruin & Celestron C11 0.28\,m & Baader Dados & 7500 & H$\alpha$ & 4 \\
        X.\ Dupont & Dall-Kirkham 0.35\,m & StarEx2400 & 15000 & H$\alpha$ & 19 \\
        A.\ Favaro & Celestron C8 0.2\,m & LHiResIII & 17000 & H$\alpha$ & 1\\
        P.\ Fricker & Celestron C11 0.28\,m & LHiResIII & 18000 & H$\alpha$ & 3 \\
        O.\ Garde & RC400 0.4\,m & Eshel & 11000 & echelle & 1 \\
        O.\ Gayrard & Sky-Watcher 80ED 0.08\,m & StarEx 2400 & 10000 & H$\alpha$ & 1 \\
        J.\ Guarro Fl\'o & Meade SC16 0.41\,m & NOU\_T & 8500 & echelle & 9\\
        J.\ Guarro Fl\'o & Meade SC16 0.41\,m & MUSSOL & 9000 & echelle & 25\\
        S.\ G\"ussregen & GSO RC10 0.25\,m & LHiResIII & 16000 & H$\alpha$ & 2\\
        F.\ Houpert & Celestron C11 0.28\,m& LHiRes \#194 & 15000 & H$\alpha$ & 4\\
        M.\ Larsson & Celestron C8 0.2\,m & StarEx2400 & 15000 & H$\alpha$ & 1\\
        R.\ Leadbeater & Celestron C11 0.28\,m & LHiRes \#29 & 17000 & H$\alpha$ & 4\\
        V.\ Lecocq & M703 0.18\,m & StarEx2400 & 15000 & H$\alpha$ & 2\\
        T.\ Lester & Dall-Kirkham 0.31\,m & long slit & 8000 & H$\alpha$ & 1\\
        R.\ Pomillo & Celestron C11 0.28\,m & StarEx2400 & 12000 & H$\alpha$ & 3\\
        O.\ Thizy & Celestron C11 0.28\,m& eShel & 11000 & echelle & 3\\
        F.\ Weil & RC200 0.2\,m & StarEx2400 & 16900 & H$\alpha$ & 1\\
        \hline
        \end{tabular}}
        \tablefoot{2SPOT stands for Southern Spectroscopic Project Observatory Team, that is S.\ Charbonnel, P.\ Le D\^u, O.\ Garde, L.\ Mulato and T.\ Petit. R is the resolving power, whilst N corresponds to the number of observations that were collected.}
    \end{table}
    
    \section{Selected H$\alpha$ line profiles \label{app1}}
\begin{figure}[h]
    \begin{center}
      \resizebox{8cm}{!}{\includegraphics{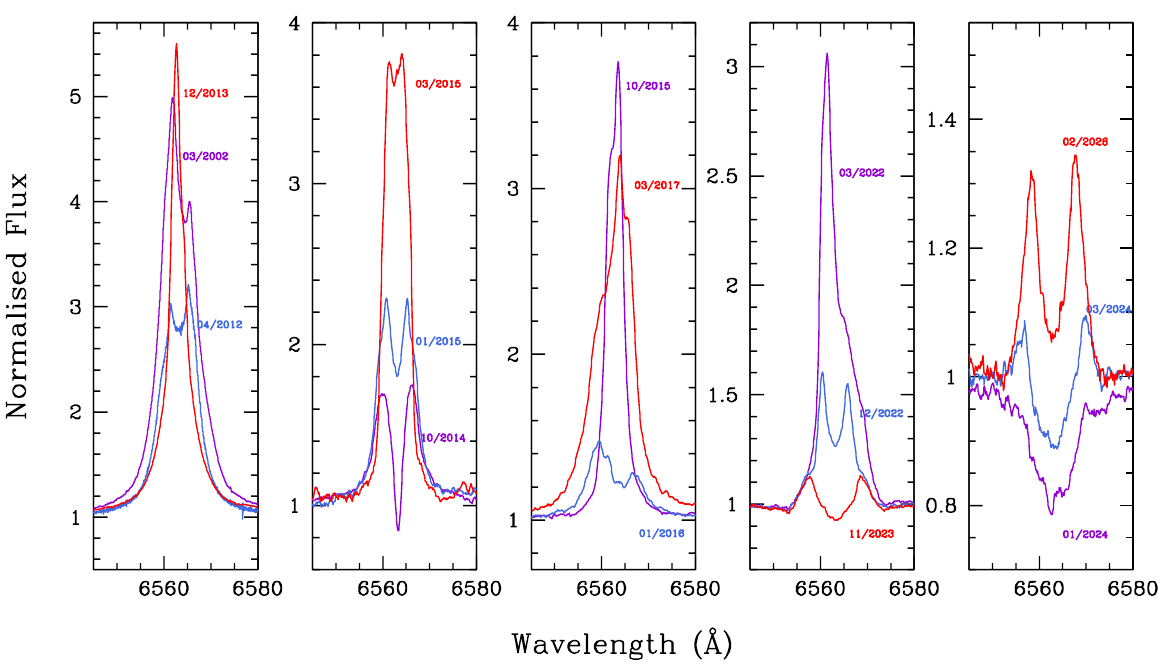}}
      \caption{Representative H$\alpha$ line profiles observed during the different stages in the long-term evolution of the disc of HD\,45314. From left to right, the different panels correspond to the high activity state, the shell episode and the subsequent return to normal double-peaked line morphology, the oscillating phase, the decline phase, and the low activity and slow recovery state. \label{montage}}
    \end{center}
\end{figure}

    Figure\,\ref{montage} illustrates some selected line profiles for each of the five stages in the long-term evolution of HD\,45314's circumstellar disc. In each panel, the oldest observation is shown in violet colour, the second oldest in blue, and the most recent one in red.  

In Fig.\,\ref{montage}, only the first of the three spectra in the second panel displays a genuine shell profile. The shell episode probably started before the beginning of the star's visibility season. Within two months after the October 2014 observation, the shell profiles gave way to normal double-peaked profiles \citep[see also Figs.\,6 and 8 of][]{Rau18}.

\begin{figure}[h]
    \begin{center}
      \resizebox{8cm}{!}{\includegraphics{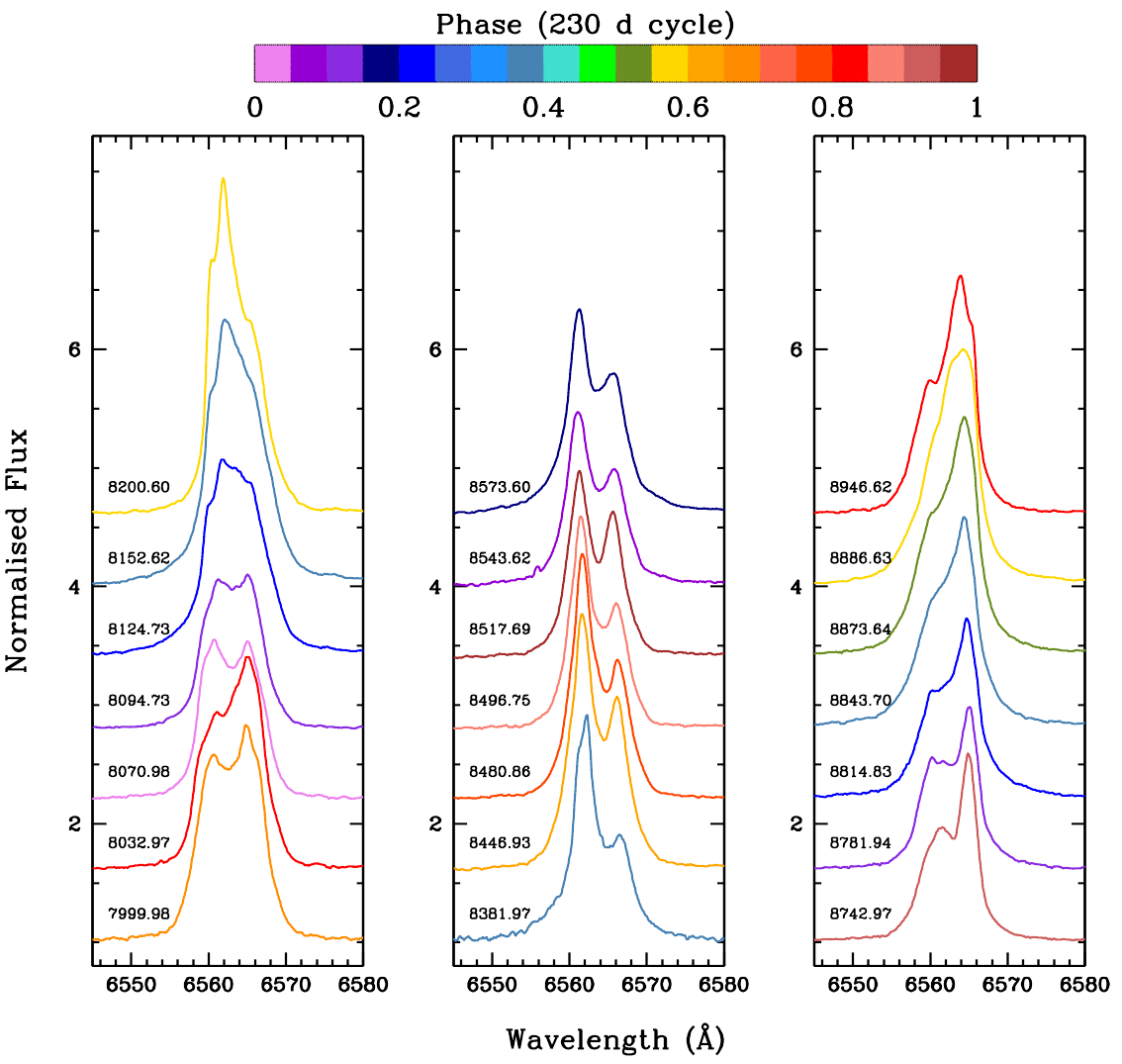}}
      \caption{Representative H$\alpha$ line profiles observed during three oscillation cycles. From left to right, data were taken during the 2017-2018, 2018-2019, and 2019-2020 observing seasons. In each panel, the spectra are shifted upwards in chronological order by 0.6 continuum units. The labels indicate the time of the observation in the format HJD $-$ 2\,450\,000. The colours correspond to the phase during the 230\,d cycle. The colour scale at the top runs from minimum EW (phase 0.0 taken at HJD\,2\,458\,070.981) to maximum (phase 0.5) and back to minimum. \label{montageoscill}}
    \end{center}
\end{figure}

Figure\,\ref{montageoscill} displays the H$\alpha$ line profiles observed during three cycles of the oscillation phase. Whilst the overall line strength varies with the 230\,d cycles, the line morphologies are also strongly affected by the V/R variations which occur on a longer time scale (see Sect.\,\ref{mediumterm}). We note the triple peak line morphology on the HJD\,2\,458\,781.94 spectrum. The same morphology was seen on all spectra collected around that date, indicating that this feature is real.

    \section{Long-term EW variations \label{app2}}
\begin{figure}[h]
    \begin{center}
      \resizebox{8cm}{!}{\includegraphics{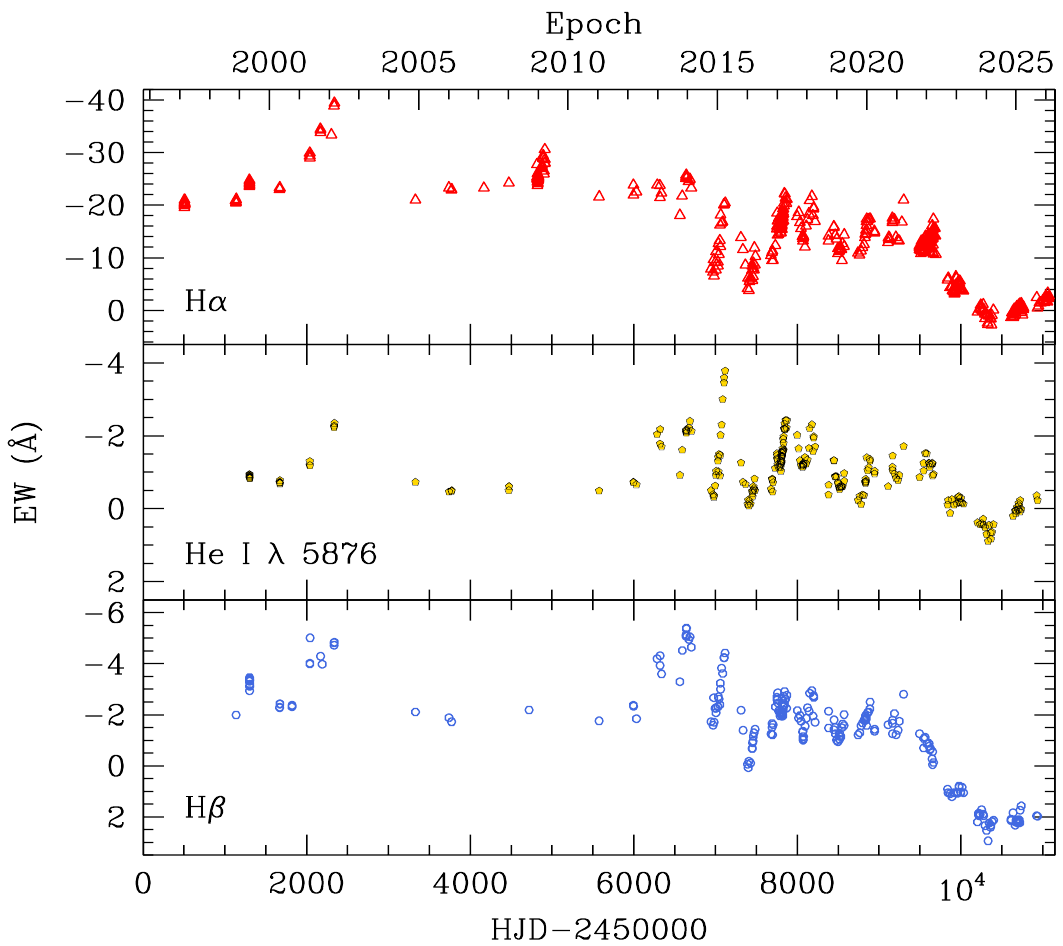}}
      \caption{Comparison of the variations in EW of the H$\beta$, He\,{\sc i} $\lambda$\,5876 and H$\alpha$ lines. \label{historicEW}}
    \end{center}
\end{figure}
\begin{figure}[h]
    \begin{center}
      \resizebox{8cm}{!}{\includegraphics{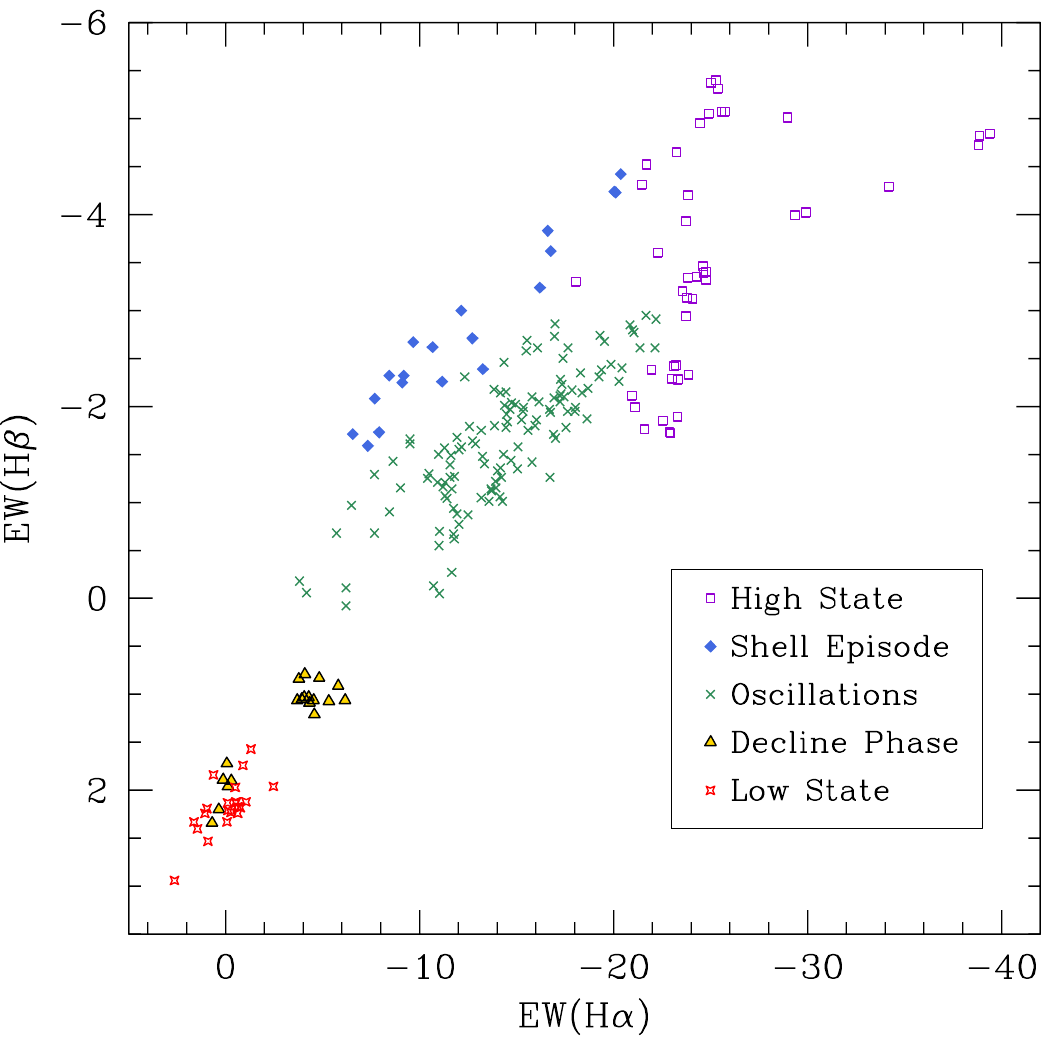}}
      \caption{EW(H$\beta$) as a function of EW(H$\alpha$) for HD\,45314. The different symbols stand for the different phases in the evolution of the circumstellar disc over the past three decades. \label{EWHaHb}}
    \end{center}
\end{figure}
\begin{figure}[h]
    \begin{center}
      \resizebox{8cm}{!}{\includegraphics{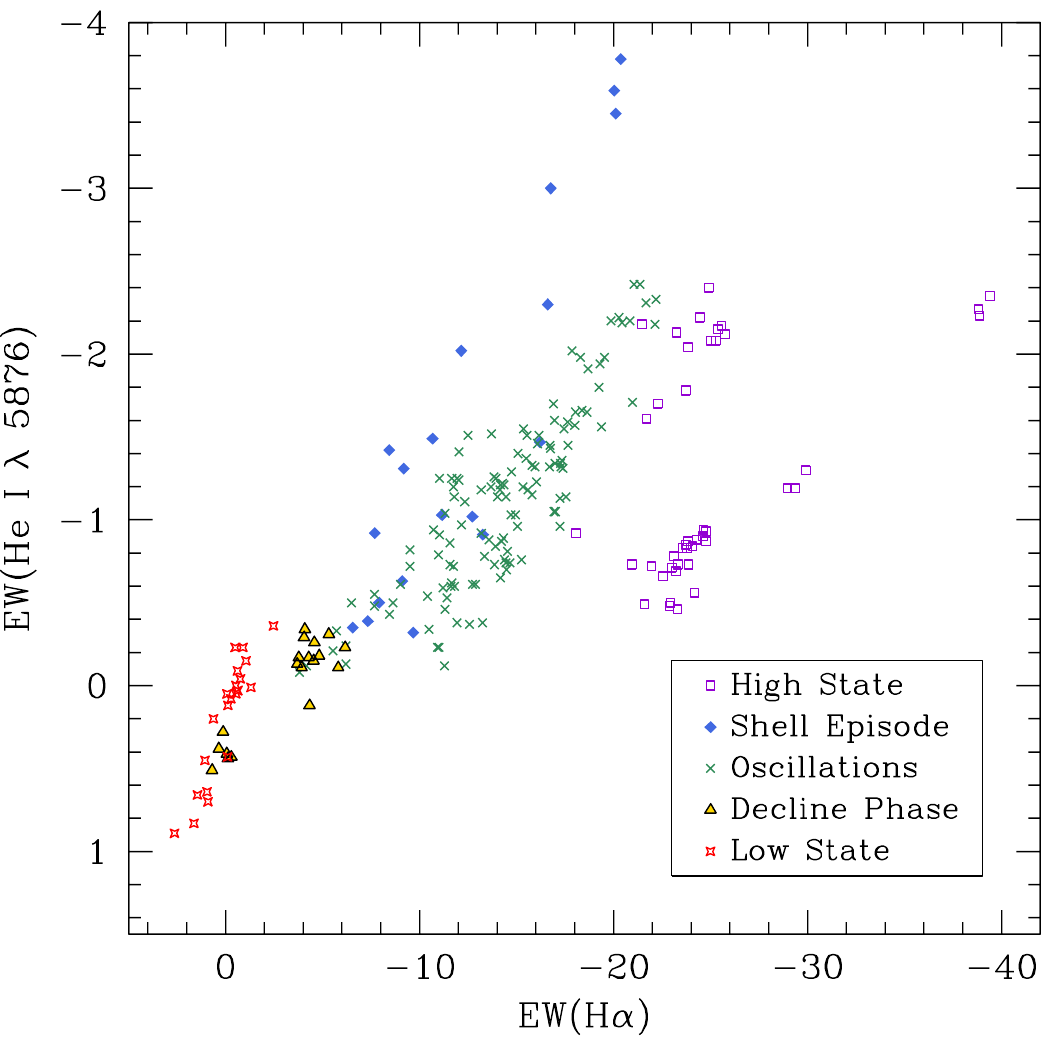}}
      \caption{Same as Fig.\,\ref{EWHaHb} but for EW(He\,{\sc i} $\lambda$\,5876) as a function of EW(H$\alpha$). \label{EWHaHe}}
    \end{center}
\end{figure}

Figure\,\ref{historicEW} illustrates the variations in EWs of the most prominent emission lines arising from HD\,45314's circumstellar disc over the past three decades. Figures\,\ref{EWHaHb} and \ref{EWHaHe} display the EWs of the H$\beta$ and  He\,{\sc i} $\lambda$\,5876 lines as a function of EW(H$\alpha$). The colours of the symbols used in Figs.\,\ref{EWHaHb} and \ref{EWHaHe} correspond to the five different states of the decretion disc introduced in Sect.\,\ref{longterm} and Fig.\,\ref{historic}. 

    \section{Minimum emission spectrum \label{app3}}
\begin{figure}[h]
    \begin{center}
      \resizebox{8cm}{!}{\includegraphics{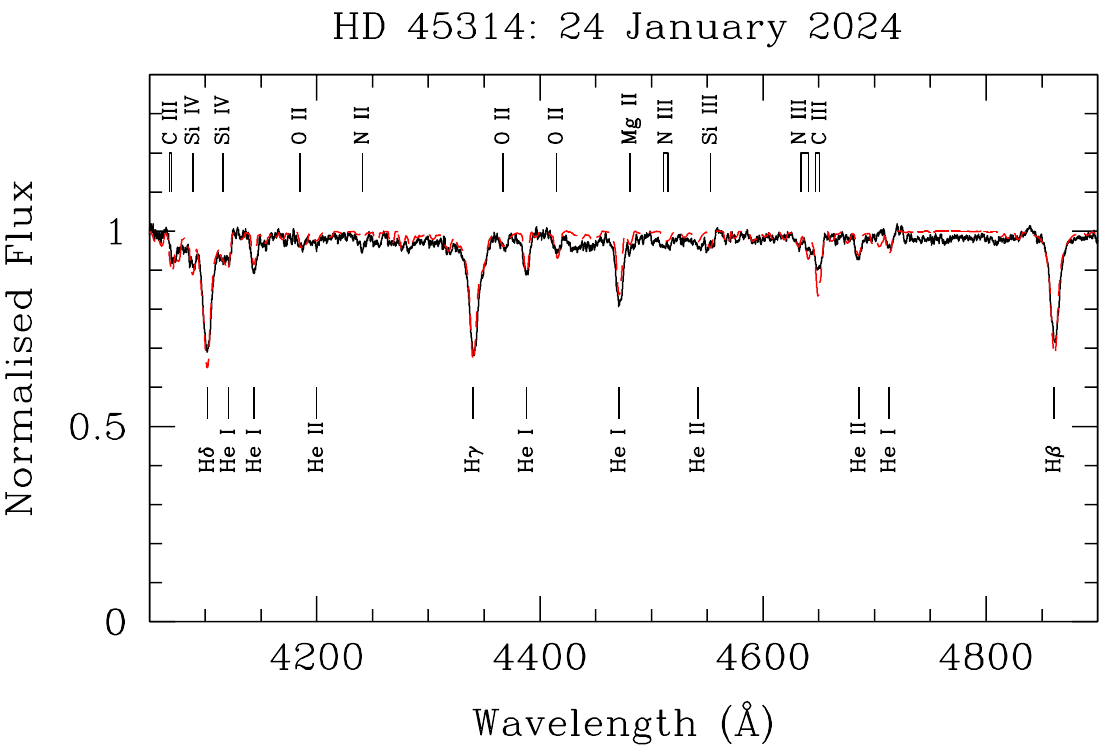}}
      \caption{Blue spectrum of HD\,45314 as observed on 24 January 2024 when the circumstellar emission reached its minimum. The red dashed line shows a TLUSTY spectrum with $T_{\rm eff} = 30$\,kK and $\log{g} = 4.0$, rotationally broadened to $v\,\sin{i} = 210$\,km\,s$^{-1}$.\label{SpT}}
    \end{center}
\end{figure}

    Figure\,\ref{SpT} illustrates the blue spectrum of HD\,45314 recorded with the TIGRE + HEROS instrument on 24 January 2024 (HJD\,2460333.883). This date corresponds to the weakest H$\alpha$ emission, though there still remained some residual emission within the H$\alpha$ line.
    
We measured the EWs of several He\,{\sc i} and He\,{\sc ii} diagnostic lines: EW(He\,{\sc i} $\lambda$\,4388) = $0.53 \pm 0.01$\,\AA, EW(He\,{\sc i} $\lambda$\,4471) = $1.12 \pm 0.04$\,\AA, EW(He\,{\sc ii} $\lambda$\,4542) = $0.13 \pm 0.01$\,\AA, EW(He\,{\sc ii} $\lambda$\,4686) = $0.31 \pm 0.01$\,\AA, and EW(He\,{\sc i} $\lambda$\,4713) = $0.20 \pm 0.01$\,\AA. Using the spectral classification criteria of \citet{Con71} and \citet{Mar18} yields a spectral type O9.7. In the same way, the luminosity class criteria of \citet{Mat89} and \citet{Mar18} suggest a giant luminosity class. The latter is however at odds with the absolute magnitude of the star that we infer below. Comparing the minimum state spectrum with the spectral atlas of \citet{Sot11} suggests a spectral type O9.7\,V -- B0\,V. Indeed, whilst the low intensity ratios of He\,{\sc ii} $\lambda$\,4542/He\,{\sc i} $\lambda$\,4388 and He\,{\sc ii} $\lambda$\,4200/He\,{\sc i} $\lambda$\,4144 strongly favour a B0 spectral type, the fact that Si\,{\sc iii} $\lambda$\,4552 and He\,{\sc ii} $\lambda$\,4542 are of comparable strengths argues instead for an O9.7 classification. Finally, comparison of the minimum state spectrum with synthetic TLUSTY spectra \citep{Lan03}, rotationally broadened to $v\,\sin{i} = 210$\,km\,s$^{-1}$, indicates an excellent agreement for $T_{\rm eff} = 30$\,kK and $\log{g} = 4.0$. The main deviations concern the cores of the H$\beta$ and H$\delta$ Balmer lines and C\,{\sc iii} $\lambda\lambda$\,4647-4650. The Balmer lines are predicted to be a few percent deeper than observed. This is likely due to residual circumstellar emission affecting those lines.

On the dates around 24 January 2024, the KWS and Vollmann data indicate $m_V = (7.14 \pm 0.02)$\,mag. The KWS photometry further yields $V-I_C = -0.006$\,mag. Adopting an intrinsic $(V-I_c)_0 \simeq -0.358$\,mag, that is the mean for an O9.5\,V and a B0\,V star according to \citet{Pec13}, we infer $E(V-I_C) = 0.352$\,mag. Using the $R_V = 3.1$ extinction law from \citet{Car89}, this leads to $A_V = 0.68$\,mag or $A_V = 0.79$\,mag if we adopt instead $R_V = 5.0$. Alternatively, we used the {\tt G-Tomo} tool\footnote{Available via https://explore-platform.eu/.} based on the highest resolution {\it Gaia}-2MASS 3D dust maps of \citet{Lal22} and \citet{Ver22} to evaluate $A_V = (0.75 \pm 0.01)$\,mag. As a conservative approach, we thus adopt the mean value of the above estimates, $A_V = (0.74 \pm 0.05)$\,mag. Based on the $(1.128 \pm 0.037)$\,mas parallax from {\it Gaia}-DR3 \citep{DR3}, \cite{Bai21} estimated a geometrical distance of $864^{+30}_{-27}$\,pc. Combining these values leads to an absolute magnitude of $M_V = (-3.28 \pm 0.09)$\,mag, which is marginally consistent with an O9.7\,V \citep{Mar05} or B0\,V \citep{Neg24} main-sequence star, but much too faint for a giant. Adopting an effective temperature of 30\,kK and a bolometric correction of $-2.86$\,mag \citep{Mar05}, we finally estimate a bolometric luminosity of 22580\,L$_{\odot}$, that is $8.68 \times 10^{37}$\,erg\,s$^{-1}$.

\section{Trends in {\it TESS} data \label{trendsTESS}}
\begin{figure}[h]
    \begin{center}
      \resizebox{8cm}{!}{\includegraphics{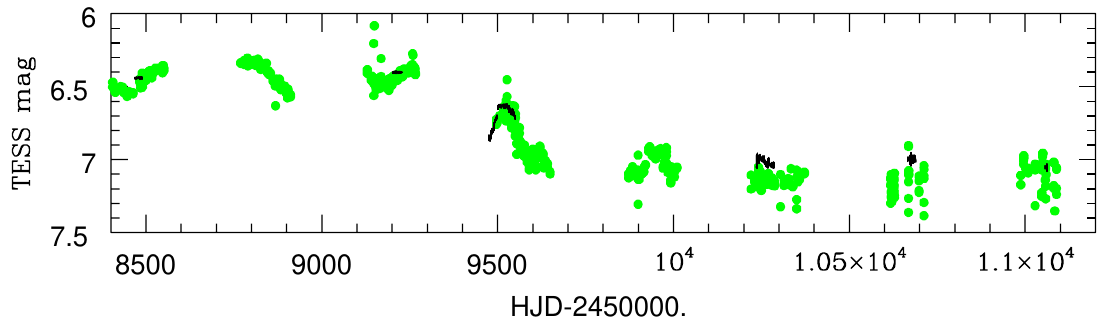}}
      \caption{Comparison of {\it TESS} PDC magnitudes (black symbols) and KWS $I_c$-band photometry (green symbols).\label{TESSvsKWS}}
    \end{center}
\end{figure}

The {\it TESS} PDC data of HD\,45314 exhibit medium-term variations in some sectors. Figure\,\ref{TESSvsKWS} compares the {\it TESS} PDC magnitudes, prior to any detrending, to the $I_c$ photometry from the KWS project. There is a rather good agreement between these data as far as the medium-term variations are concerned.

\begin{figure}[h]
    \begin{center}
      \resizebox{8cm}{!}{\includegraphics{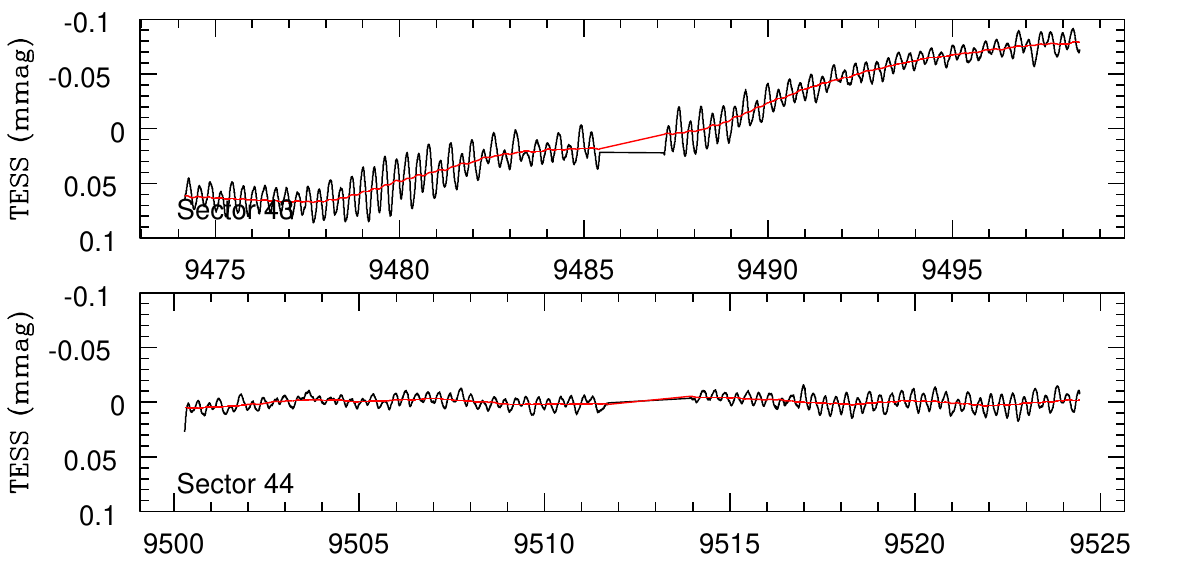}}
      \resizebox{8cm}{!}{\includegraphics{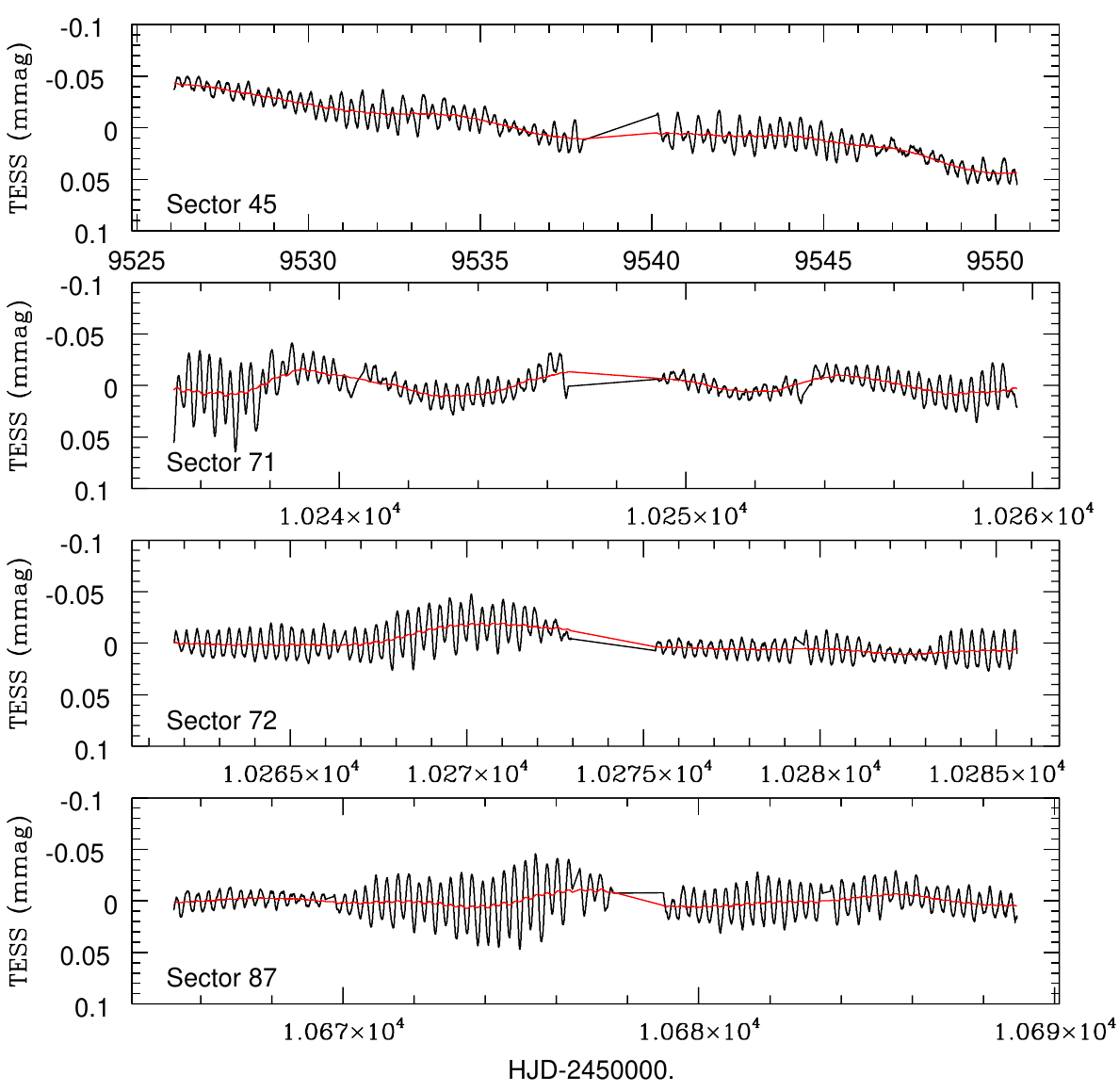}}
      \caption{{\it TESS} light curves before detrending (black data) and derived trends (red curve). From top to bottom, the panels correspond to Sectors\,43, 44, 45, 71, 72, and 87.\label{detrending}}
    \end{center}
\end{figure}

Those sectors displaying significant medium-term trends were detrended as explained in Sect.\,\ref{short}. Figure\,\ref{detrending} displays the PDC data before detrending along with the trends that were subtracted prior to the Fourier analysis. 

\section{Short-term spectroscopic variations \label{appshort}}
To assess the significance of short-term variations in the optical spectrum of HD\,45314, we computed the temporal variance spectrum \citep[TVS,][]{Ful96} of the time series of spectra from our intensive monitoring with TIGRE (December 2019) and Aur\'elie (September 2020 and October 2021).

\begin{figure*}[h]
  \begin{minipage}{5.5cm}
    \resizebox{5.5cm}{!}{\includegraphics{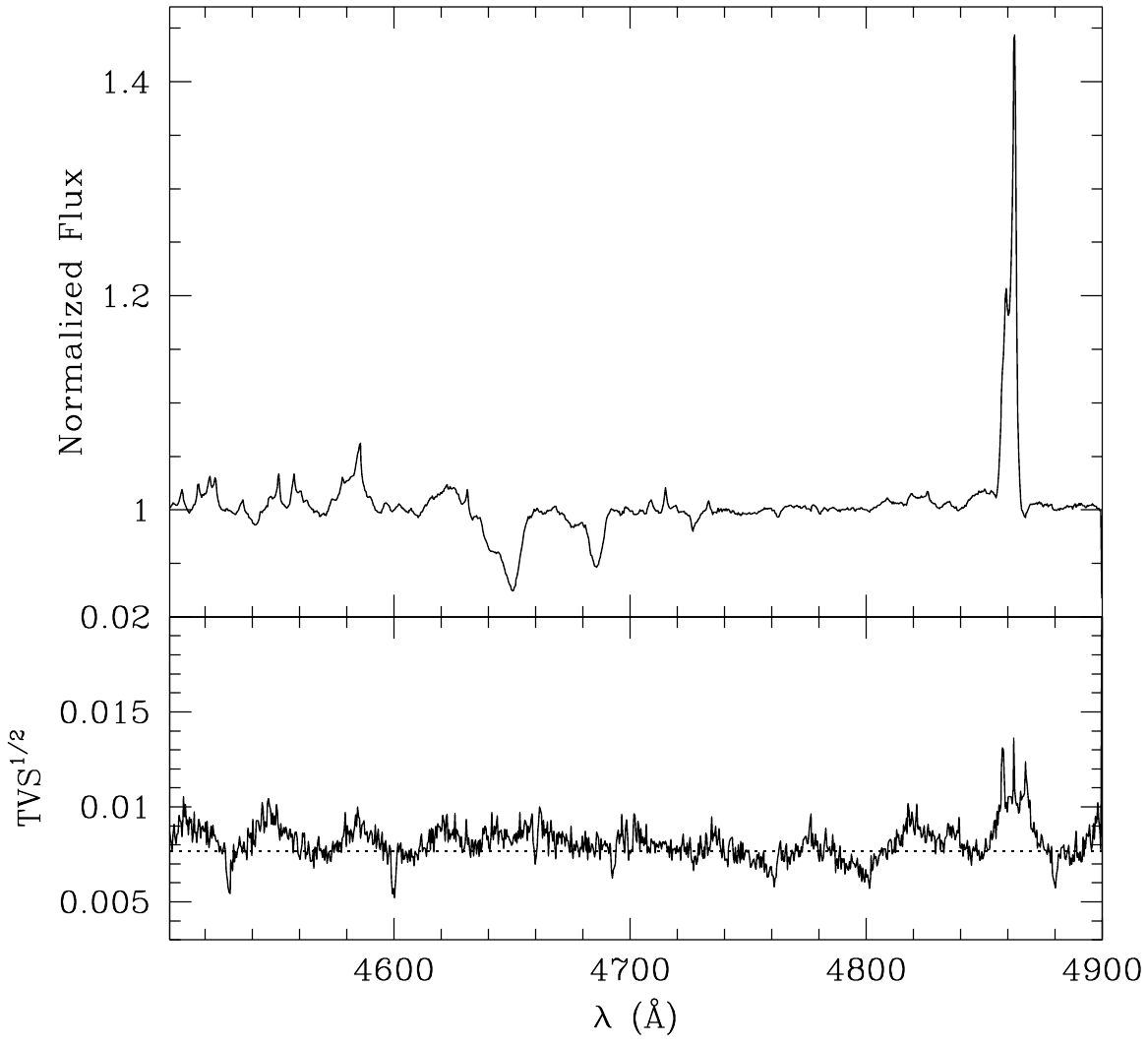}}
  \end{minipage}
  \hfill
  \begin{minipage}{5.5cm}
    \resizebox{5.5cm}{!}{\includegraphics{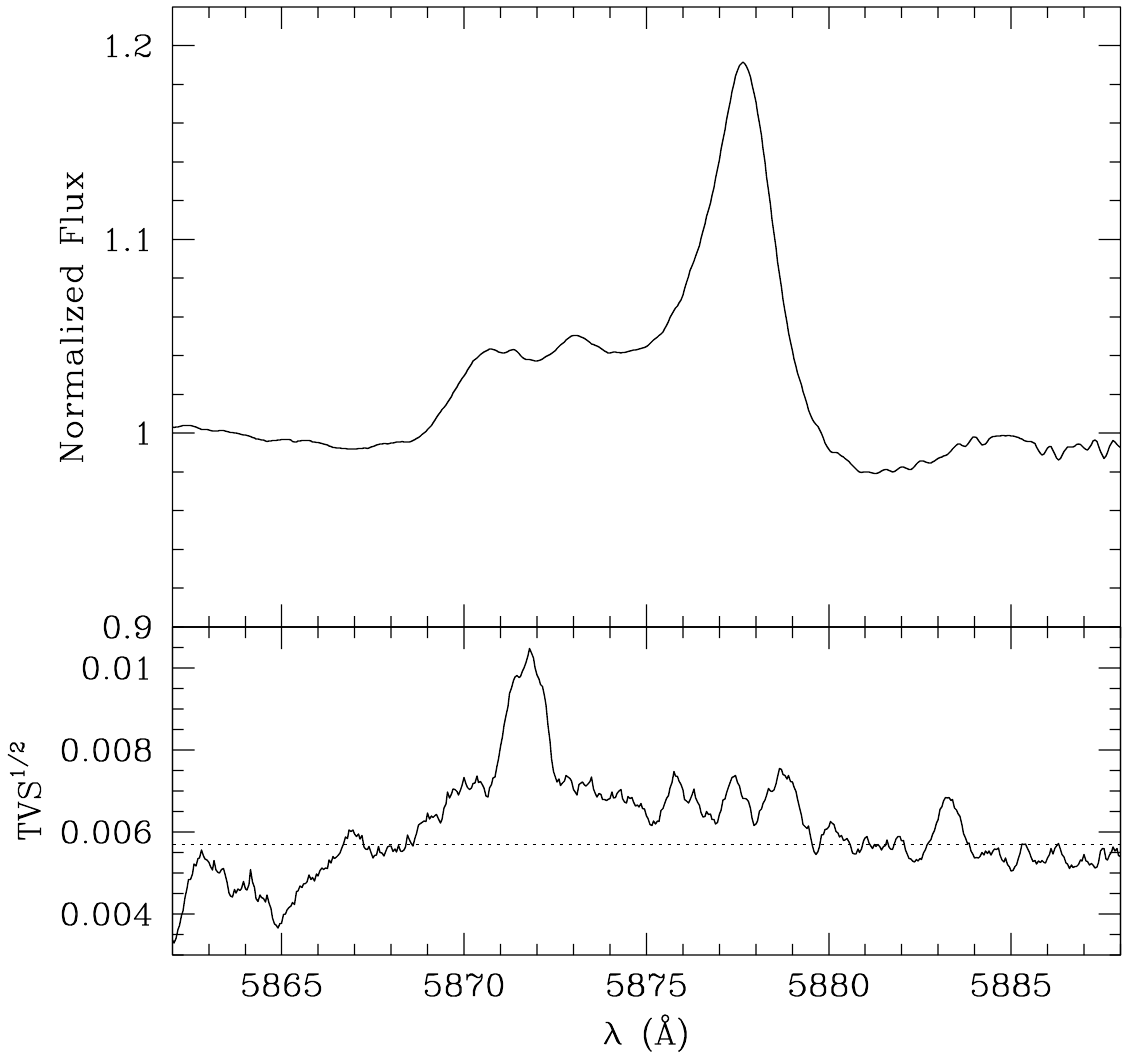}}
  \end{minipage}
  \hfill
  \begin{minipage}{5.5cm}
    \resizebox{5.5cm}{!}{\includegraphics{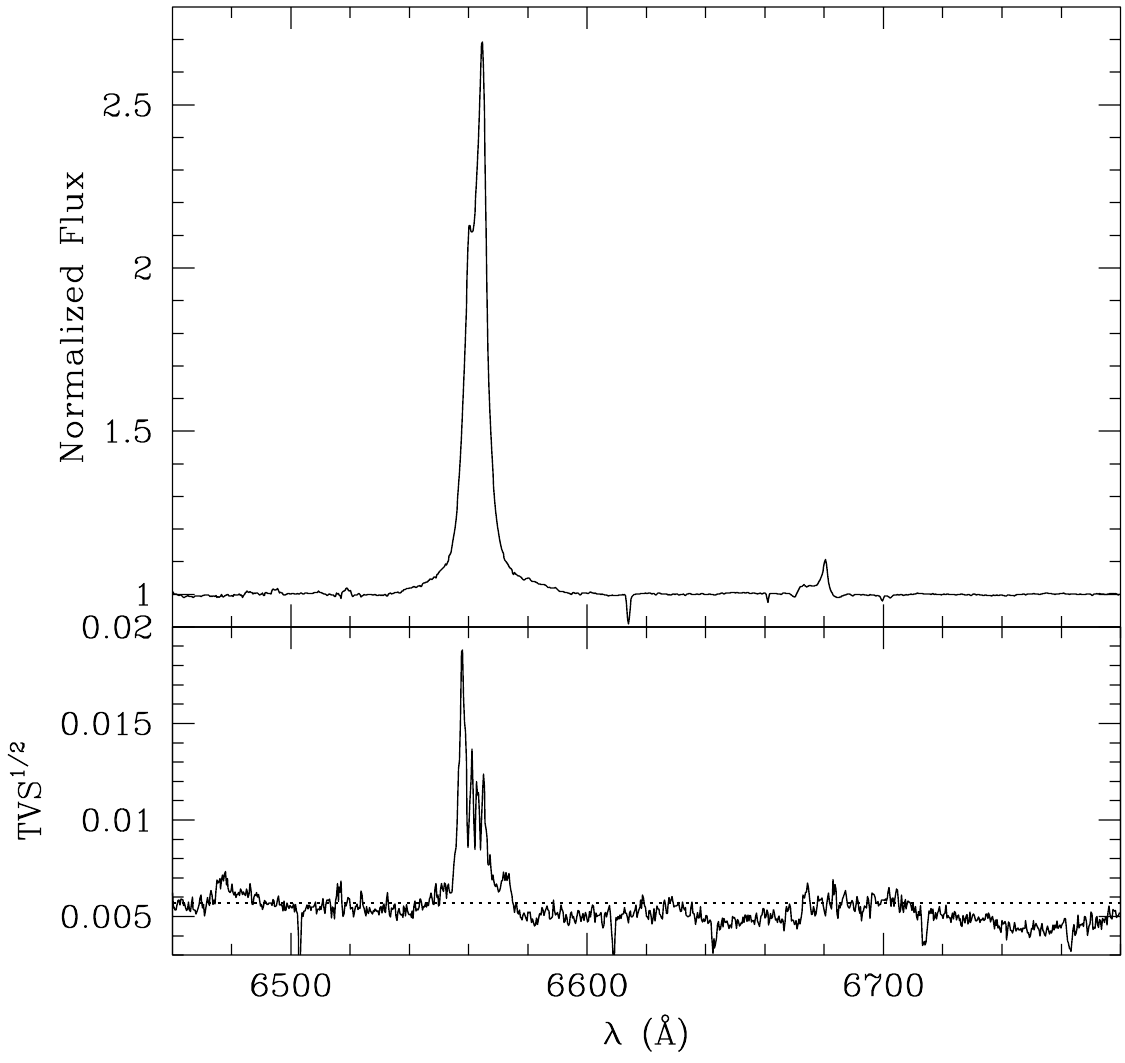}}
  \end{minipage}
  \begin{minipage}{5.5cm}
    \resizebox{5.5cm}{!}{\includegraphics{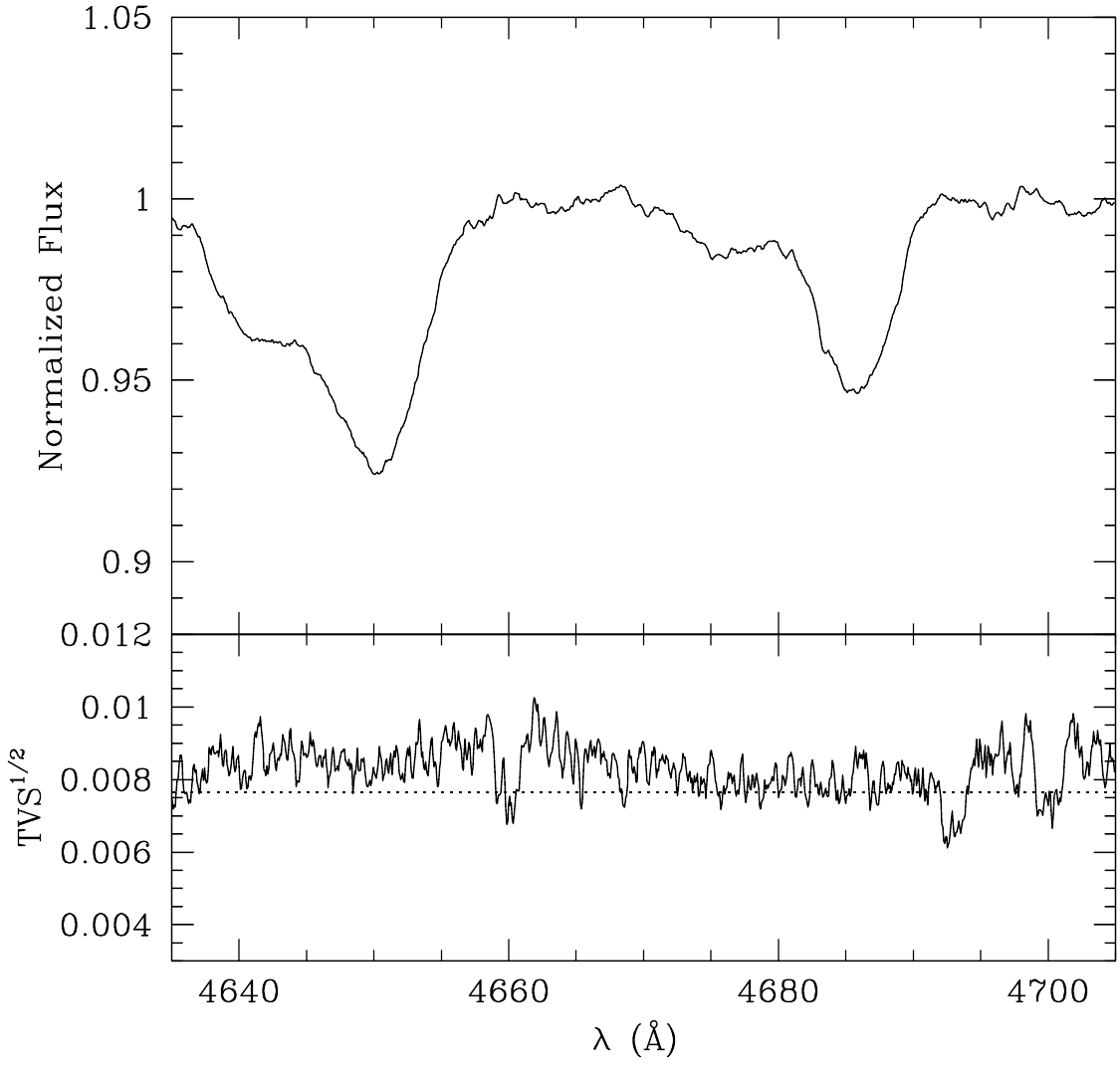}}
  \end{minipage}
  \hfill
  \begin{minipage}{5.5cm}
    \resizebox{5.5cm}{!}{\includegraphics{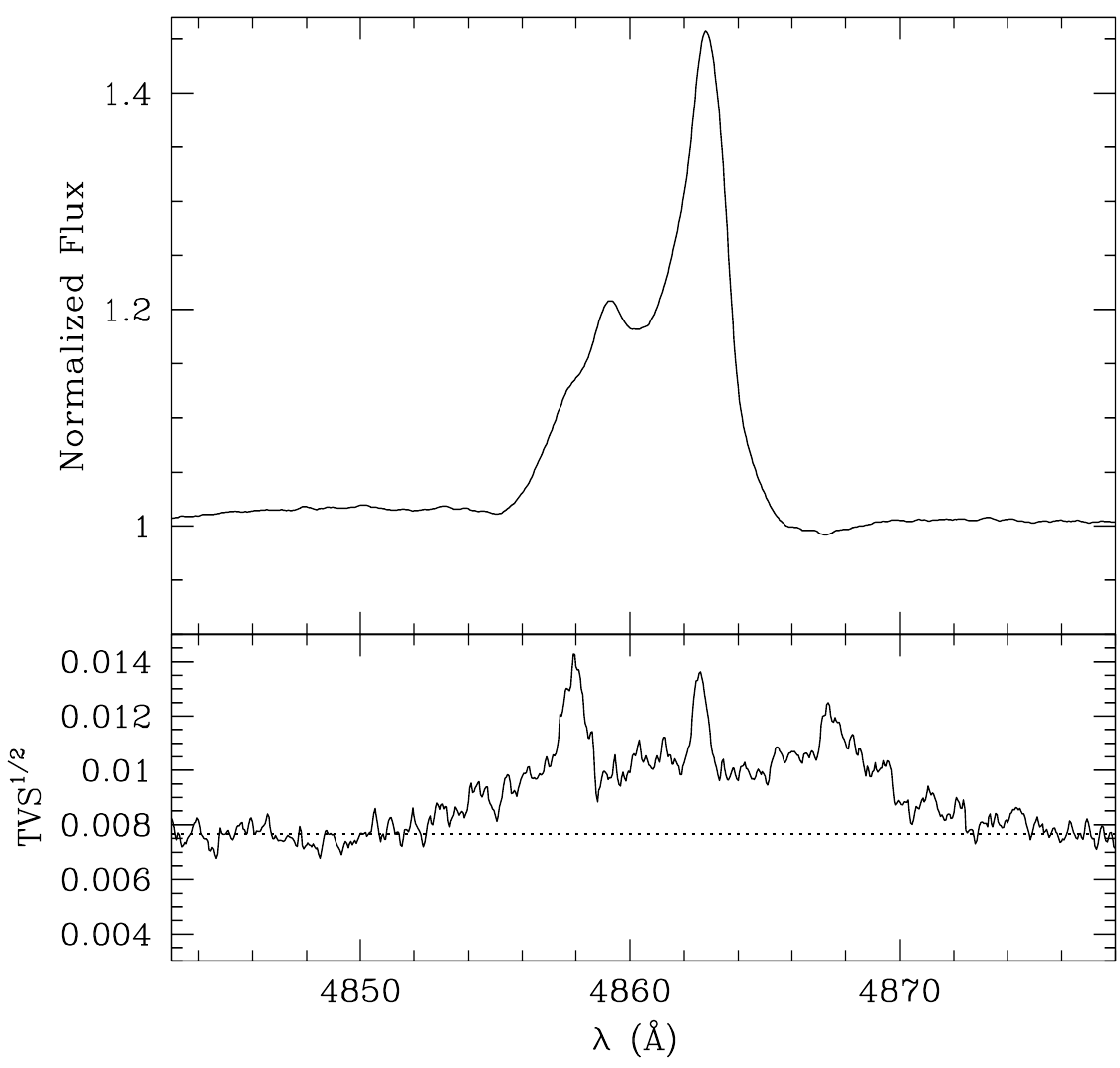}}
  \end{minipage}
  \hfill
  \begin{minipage}{5.5cm}
    \resizebox{5.5cm}{!}{\includegraphics{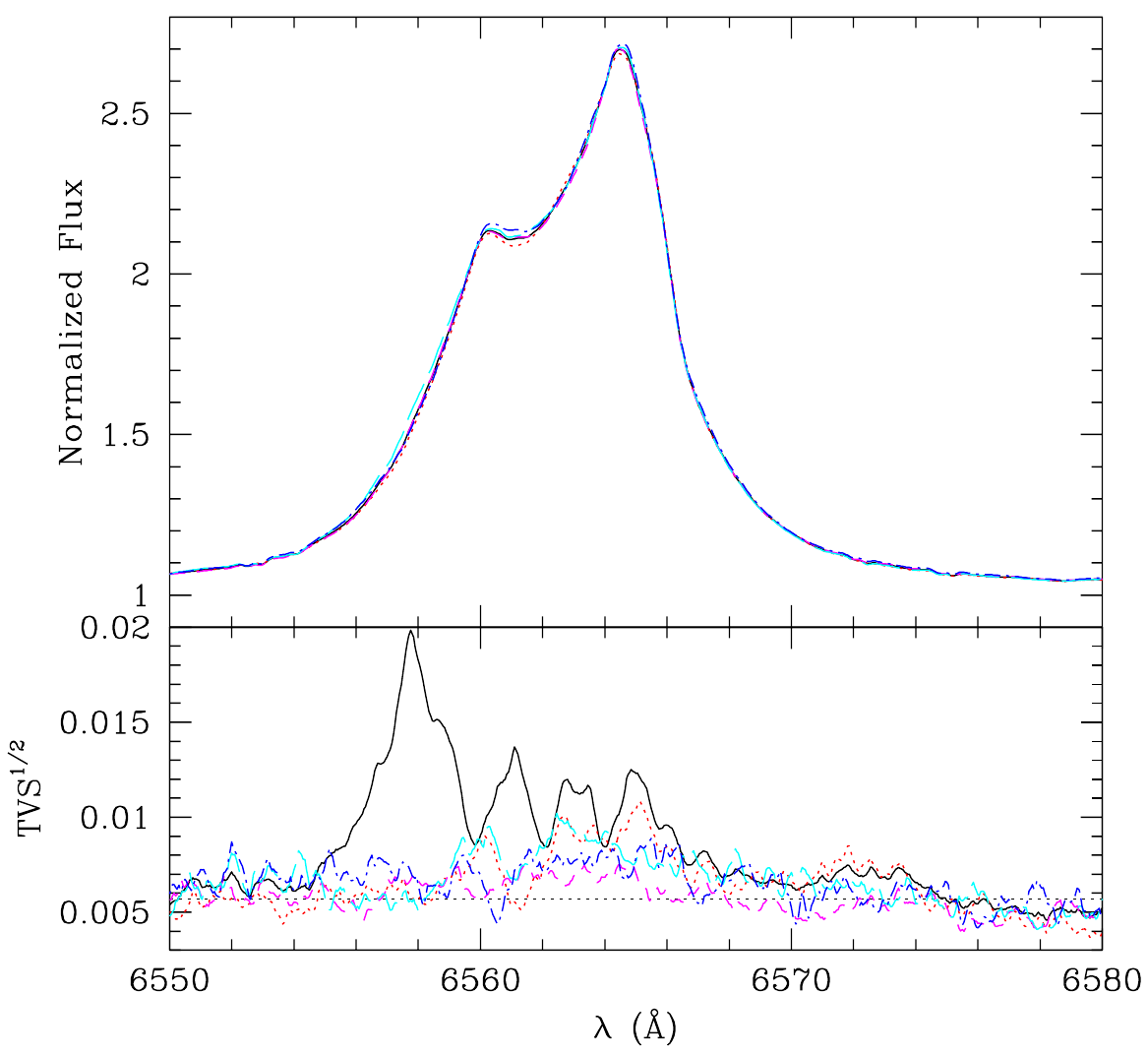}}
  \end{minipage} 
          \caption{Top row: mean spectra and $TVS^{1/2}$ computed from the TIGRE intensive monitoring campaign. The dotted line in the $TVS^{1/2}$ panel indicates the 99\% significance level estimated from the signal-to-noise ratio of the data. Bottom row: zoom onto the blue region used for RV determination via cross-correlation (left), the H$\beta$ (middle) and H$\alpha$ (right) emission lines. In the bottom right panel, the black solid line corresponds to the mean profile and $TVS^{1/2}$ of the full time series, whilst the red, magenta, cyan and blue curves show the results of the intra-night variability study respectively for the first, second, third and fourth night of the intensive TIGRE monitoring campaign.\label{TVS_TIGRE}}
\end{figure*}

For TIGRE, we studied several spectral regions, from 4500 to 4900\,\AA, from 5860 to 5890\,\AA, and from 6450 to 6790\,\AA. Prominent emission lines (H$\beta$, He\,{\sc i} $\lambda$\,5876, and H$\alpha$) exhibit line profile variability well above the 99\% significance level. An intriguing behaviour is observed for the strongest peak in the $TVS^{1/2}$ of H$\alpha$, near 6558\,\AA. This peak is absent from the $TVS^{1/2}$ computed with data from individual nights of observations (see bottom right panel of Fig.\,\ref{TVS_TIGRE}). Whilst the intra-night variability remains significant, it is of much lower amplitude than the variability over the full campaign. Comparing the mean line profiles from the four nights, we find that the strongest peak actually comes from a bump in the blue wing of the emission line that was present during the third night. In a similar way, the second strongest peak of the global $TVS^{1/2}$ near 6561\,\AA\ does not have a clear counterpart in the intra-night variations. This latter feature arises from inter-night variations in the plateau between the violet and red peak of the emission. Therefore, the strongest variability seen in the prominent emission lines probably arises on timescales of a few days and might thus stem from density variations in the disc. Such variations could be similar to those found by \citet{Lab25} in a sample of Be stars during mass injection events.

\begin{figure*}[h]
  \begin{minipage}{5.5cm}
    \resizebox{5.5cm}{!}{\includegraphics{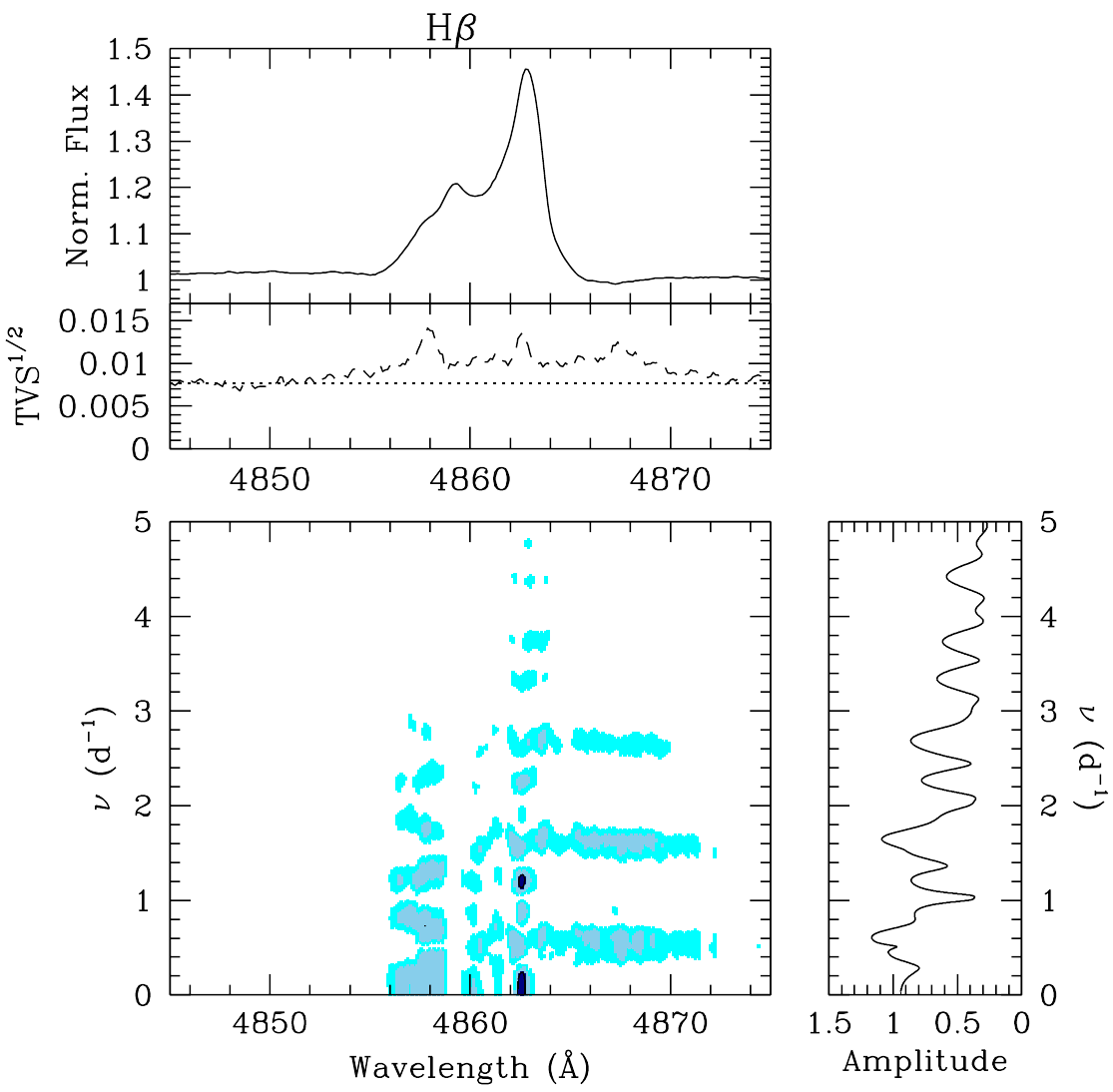}}
  \end{minipage}
  \hfill
  \begin{minipage}{5.5cm}
    \resizebox{5.5cm}{!}{\includegraphics{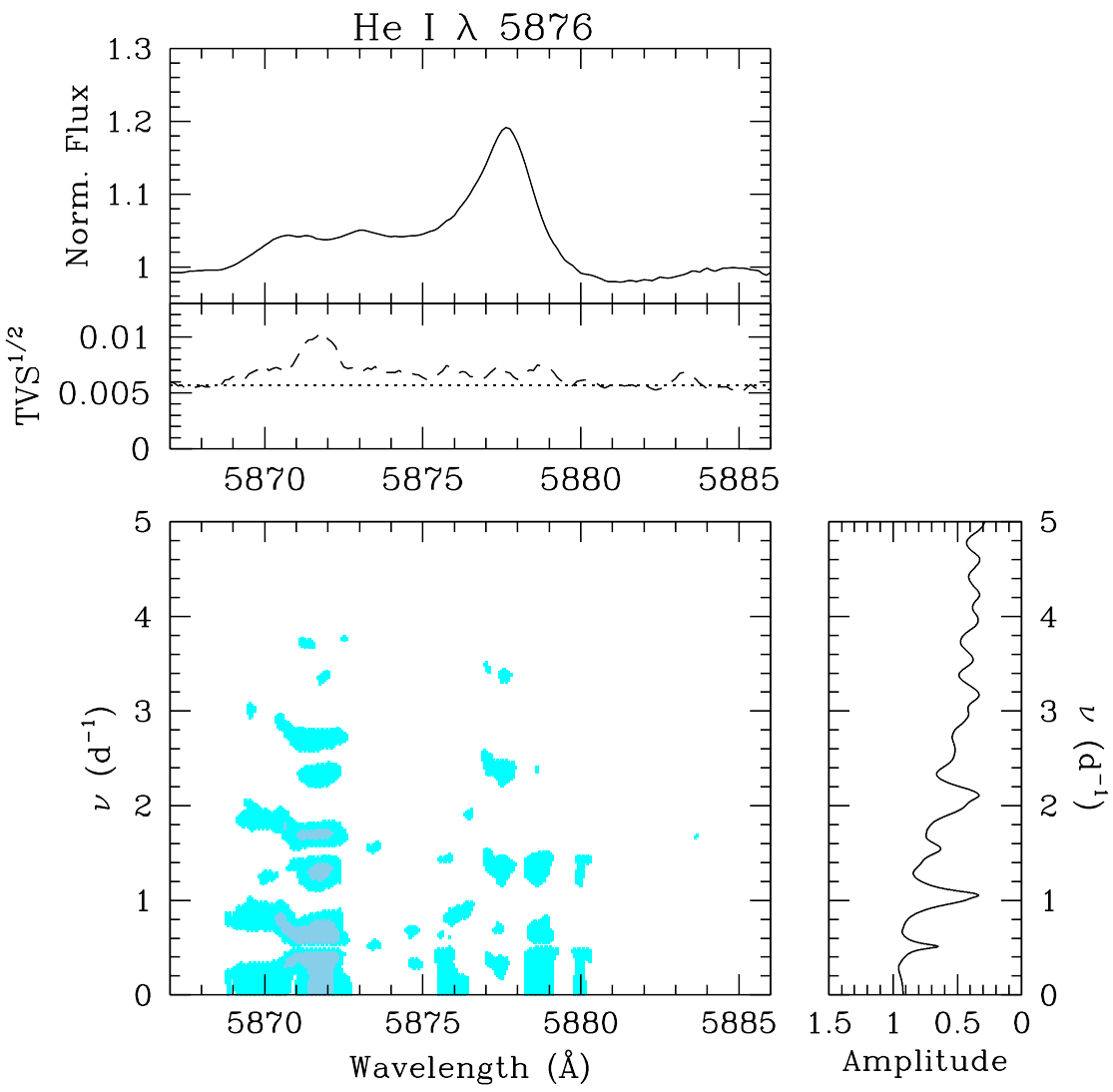}}
  \end{minipage}
  \hfill
  \begin{minipage}{5.5cm}
    \resizebox{5.5cm}{!}{\includegraphics{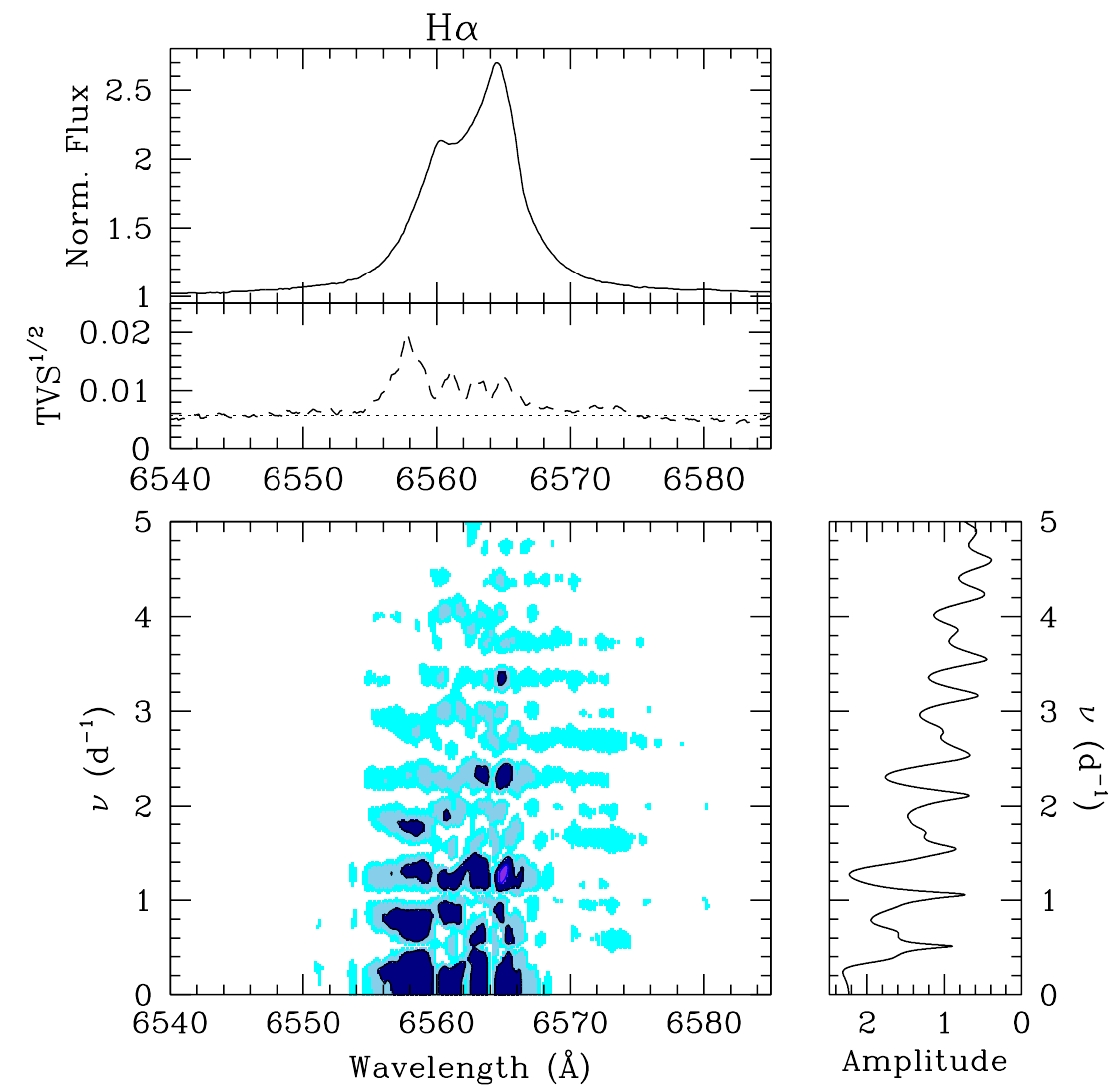}}
  \end{minipage}
  \caption{Fourier periodograms of the variations in the H$\beta$, He\,{\sc i} $\lambda$\,5876, and H$\alpha$ lines (from left to right) during the intensive TIGRE monitoring campaign. Each group of subpanels yields the mean line profile and $TVS^{1/2}$ (top panels), the Fourier periodogram normalised with respect to the 1\% significance level as a function of wavelength (colour scale plot), and the mean periodogram normalised with respect to the 1\% significance level averaged over the region where significant variability is detected (bottom right). The colours in the Fourier periodogram indicate ratios over the 1\% significance level of 1.0 (cyan), 1.5 (medium blue), 2.0 (dark blue), and 3.0 (violet). The wavelength ranges used to evaluate the mean periodogram are 4854 -- 4870\,\AA, 5868 -- 5880\,\AA, and 6556 -- 6566\,\AA, respectively for H$\beta$, He\,{\sc i} $\lambda$\,5876, and H$\alpha$.\label{Fourier_TIGRE}}
\end{figure*}

The 4500 -- 4900\,\AA\ waveband, which contains the spectral region used to search for RV variations via cross-correlation, exhibits a $TVS^{1/2}$ level above the 99\% significance threshold, although it is much lower than in the emission lines. This spectral region is dominated by several photospheric absorption lines notably N\,{\sc iii} $\lambda$\,4641, C\,{\sc iii} $\lambda\lambda$\,4647, 4650, O\,{\sc ii} $\lambda\lambda$\,4639, 4651, 4662, 4674, 4676, 4699, 4705, and He\,{\sc ii} $\lambda$\,4686. Furthermore, the US National Institute of Standards and Technology (NIST) atomic spectra database\footnote{https://physics.nist.gov/PhysRefData/ASD/lines\_form.html} lists also many Fe\,{\sc ii} lines over this spectral range. The latter could produce some weak circumstellar emission lines, so that the spectrum is probably not purely photospheric over this spectral region. 

\begin{figure}[h]
    \begin{center}
          \resizebox{8.5cm}{!}{\includegraphics{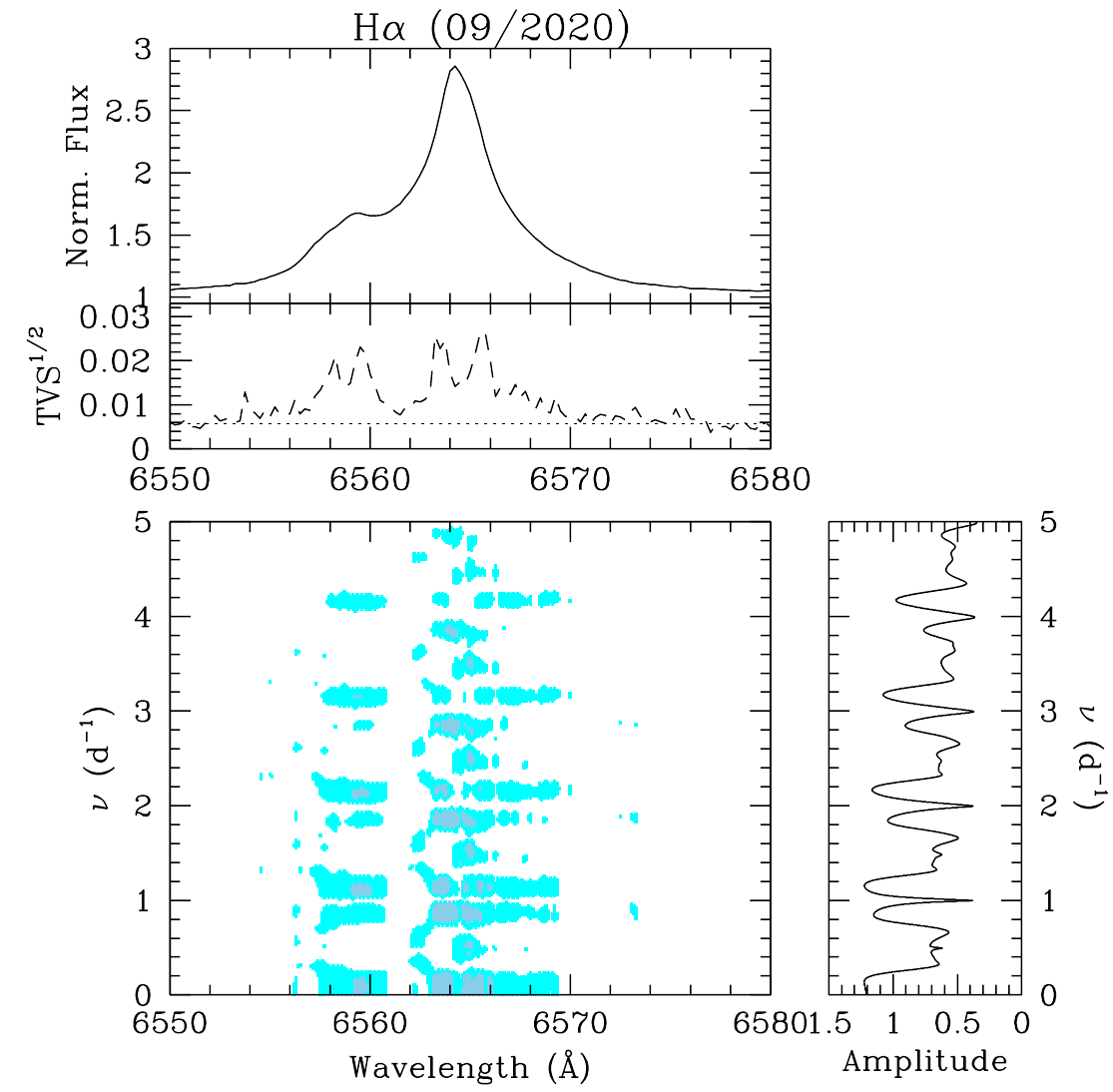}}
          \resizebox{8.5cm}{!}{\includegraphics{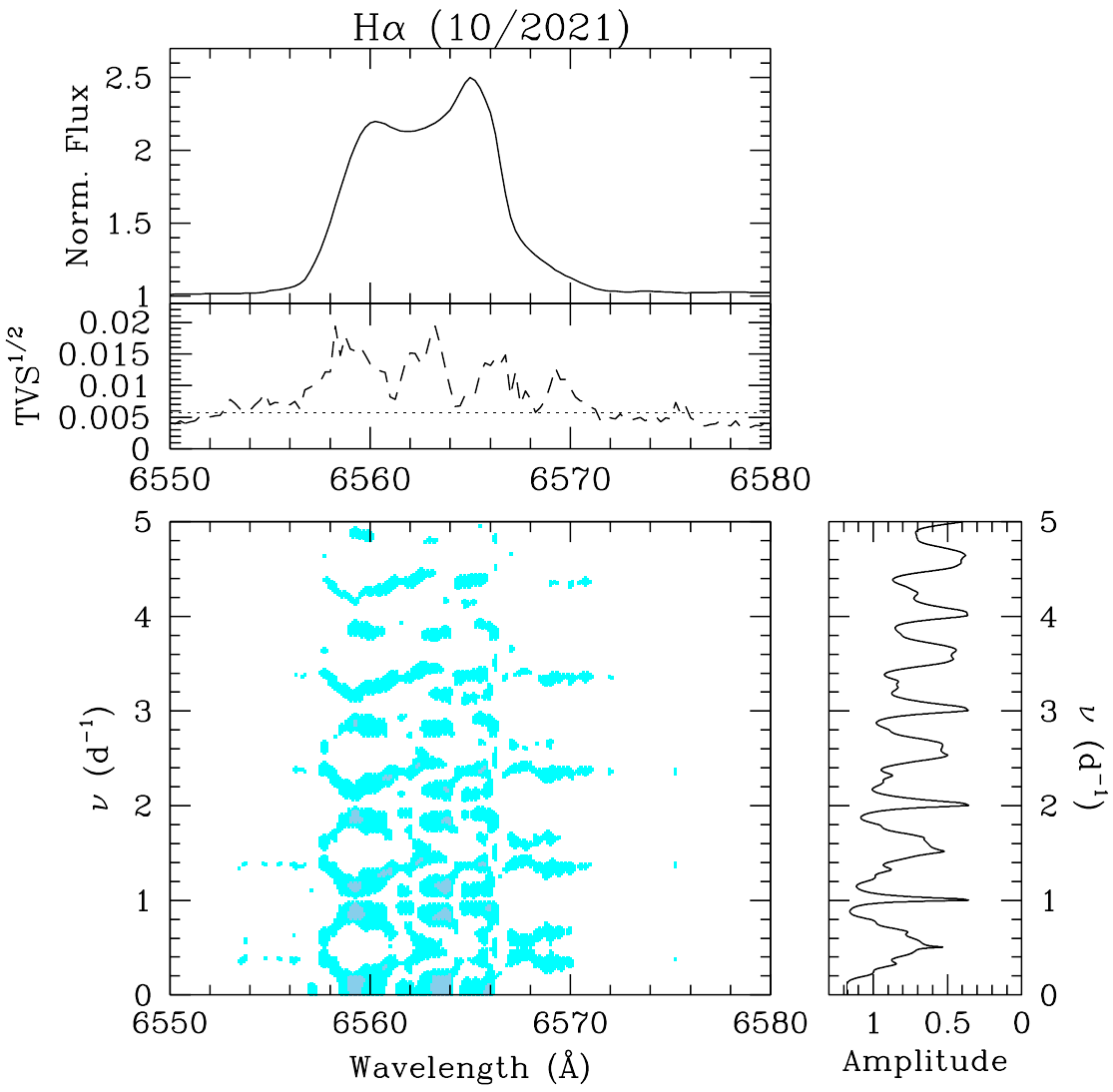}}
          \caption{Same as Fig.\,\ref{Fourier_TIGRE} for the spectral region near H$\alpha$ observed with the Aur\'elie spectrograph in September 2020 (top) and October 2021 (bottom). \label{Aurelie_intense}}
    \end{center}
\end{figure}

We built the Fourier spectra of the spectroscopic variations to search for hints of short-term periodicities. To assess the significance level of peaks in the Fourier spectra, we applied a bootstrapping method, where the pairs of times and normalised fluxes at a specific wavelength were mixed randomly. Each reshuffled artificial timeseries was analysed with the same Fourier method. This process was repeated a thousand times. For each realisation the amplitude of the highest peak in the Fourier periodogram was recorded. The histogram of these amplitudes was then used to determine a threshold value which is such that 99\% of the amplitudes of the reshuffling simulations fall below it. This procedure was applied to timeseries of normalised fluxes extracted at six different wavelengths sampling different levels of the $TVS^{1/2}$ around the H$\alpha$ line, as well as six wavelengths around H$\beta$. This enabled us to derive a simple linear scaling relation between $TVS^{1/2}$ and the 1\% significance level of the Fourier spectrum.    

The only spectral lines that show peaks in their Fourier spectra exceeding the 1\% significance level are the H$\beta$, He\,{\sc i} $\lambda$\,5876 and H$\alpha$ emission lines (see Fig.\,\ref{Fourier_TIGRE}). In each case, those peaks occur at frequencies below 5\,d$^{-1}$. For the H$\alpha$ lines, the periodogram is dominated by $0.237$\,d$^{-1}$ and its $\nu + 1$\,d$^{-1}$ and $\nu + 2$\,d$^{-1}$ aliases. Since this frequency corresponds to less than the inverse of the duration of our campaign, it is unlikely to indicate a genuine periodicity but rather reflects the longer term variability of the disc emission. The situation is very similar for the He\,{\sc i} $\lambda$\,5876 line, which has its strongest peak at $0.288$\,d$^{-1}$. The case of H$\beta$ is different insofar that the strongest peak occurs at $0.606$\,d$^{-1}$.

Very similar results were obtained for the intensive monitoring carried out with the Aur\'elie spectrograph in September 2020 and October 2021 (see Fig.\,\ref{Aurelie_intense}). In September 2020, the Fourier periodogram of the H$\alpha$ line profile variations yields the strongest peak at 0.127\,d$^{-1}$, whilst it occurs at 0.100\,d$^{-1}$ in October 2021. These results corroborate our conclusion that the observed variability stems from longer-term trends rather than from genuine periodicities. In particular, the main photometric signals at $\nu_1 = 3.369$\,d$^{-1}$ and $10.480$\,d$^{-1}$ are not detected, despite a sampling allowing their detection.

\begin{figure}[h]
    \begin{center}
          \resizebox{8.5cm}{!}{\includegraphics{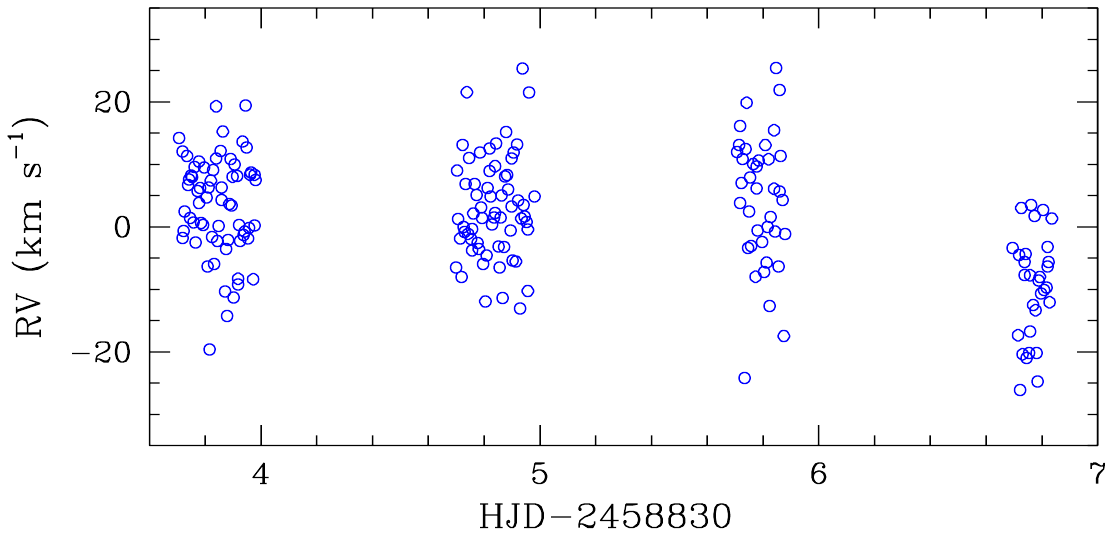}}
      \caption{RVs of HD\,45314 as determined from the 4635 - 4705\,\AA\ spectral range via cross-correlation with a synthetic TLUSTY spectrum for the four nights of intensive monitoring with TIGRE in December 2019. Only spectra with a S/N ratio better than 100 were retained. The intra-night 1\,$\sigma$ dispersions of the data amount to 7.3, 8.1, 10.5, and 7.8\,km\,s$^{-1}$ respectively for the first, second, third, and fourth night. Typical uncertainties of individual RV measurements are of order 5\,km\,s$^{-1}$. \label{RVintense}}
    \end{center}
\end{figure}

The RVs obtained via cross-correlation of the TIGRE spectra with a synthetic TLUSTY template \citep{Lan03} over the 4635 -- 4705\,\AA\ spectral range are displayed in Fig.\,\ref{RVintense}. The RVs vary quite significantly and in a non-deterministic way during each of the four nights. The intra-night dispersion amounts to 8.4\,km\,s$^{-1}$, whereas the dispersion of the complete dataset is 9.5\,km\,s$^{-1}$. This comparison suggests that the dispersion of the RVs collected over a long timescale actually results from stochastic variations such as seen during the intensive TIGRE monitoring.    

\section{Additional X-ray data \label{appXother}}
        Beside the data described in Sect.\,\ref{Xraydata} and analysed in Sect.\,\ref{SectX}, HD\,45314 was also observed by a few other X-ray satellites. \citet{Pet82} reported a possible detection with {\it UHURU} at a count rate of $1.1 \pm 0.4$\,ct\,s$^{-1}$, corresponding to a flux of $\sim 1.8 \times 10^{-11}$\,erg\,cm$^{-2}$\,s$^{-1}$ in the 2 -- 6\,keV band. This latter value would be significantly (by a factor 32) higher than measured on our April 2012 high-state observation. We stress though that this result must be taken with caution, as the detection was only at the 2.6\,$\sigma$ level, and source confusion remains a possibility given the limited angular resolution of the {\it UHURU} instrument.       
        The {\it EINSTEIN} Imaging Proportional Counter (IPC) observed HD\,45314 for 2.2\,ks on 1 April 1980. A count rate of $\leq 0.0125$\,ct\,s$^{-1}$ was inferred by \citet{Chl89}. This limit is consistent with the expected IPC count rate (0.0140\,ct\,s$^{-1}$) obtained by folding the high-state spectral model through the IPC response matrix.

        A 12.9\,ks {\it EXOSAT} exposure on 19 October 1984 resulted in a detection at a count rate of $0.13 \pm 0.05$\,ct\,s$^{-1}$ with the Medium Energy (ME) instrument. Folding the high-state spectral model through the ME response, we find an expected count rate of 0.070\,ct\,s$^{-1}$, which is at $1\,\sigma$ of the observed value. The {\it EINSTEIN} and {\it EXOSAT} results indicate that HD\,45314 was very likely in a high state in the early 1980s.
        
        The {\it Neil Gehrels Swift Observatory} observed HD\,45314 on 11 and 20 October 2021, respectively for 5.0 and 4.6\,ks. The {\it Swift} X-ray Telescope (XRT) was operated in Windowed Timing (WT) mode. This mode allows avoiding optical loading of the XRT CCD detectors despite the target being rather optically bright ($m_V \sim  7$). However, because of the strong background, the WT mode is optimal only for X-ray bright targets, which was not the case for HD\,45314 at the time of these observations. We nevertheless retrieved the data from the archive and processed them with the online tool\footnote{\tt https://www.swift.ac.uk/user\_objects/} hosted by University of Leicester \citep{Eva09}. The data yielded upper limits on the count rate of $1.01 \times 10^{-2}$\,ct\,s$^{-1}$ and $0.73 \times 10^{-2}$\,ct\,s$^{-1}$. Folding our best-fit low-state model of the March 2025 {\it XMM-Newton} spectra through the {\it Swift}-XRT response matrix yields a predicted XRT count rate of $0.28 \times 10^{-2}$\,ct\,s$^{-1}$, consistent with these upper limits. Assuming instead the high-state spectral model yields a predicted count rate of $2.10 \times 10^{-2}$\,ct\,s$^{-1}$, that is well above the limits. The {\it Swift}-XRT spectrum, extracted from the combined data of 11 October and 20 October yields an observed, background-corrected, flux of $1.4^{+1.1}_{-1.4} \times 10^{-13}$\,erg\,cm$^{-2}$\,s$^{-1}$ in the 0.5 - 10\,keV band. All these results are consistent with the contemporaneous {\it eROSITA} data, which indicated that HD\,45314 was in a low state in October 2021 (see Sect.\,\ref{SectX}).
  \end{appendix}
\end{document}